\documentclass[11pt,preprint,flushrt]{aastex631}
\usepackage{natbib,graphicx,amsmath,subfigure,color}
\usepackage{verbatim}
\usepackage{soul}
\usepackage{gensymb}
\usepackage{longtable}
\newcommand{\eqq}[1]{Equation~(\ref{#1})}

\newcommand{\change}[1]{}

\newcommand{\ie}{\textit{i.e.\/}}

\begin{document}

\title{An automated occultation network for gravitational mapping of the trans-neptunian solar system}

\author[0000-0001-6299-2445]{Daniel C. H. Gomes}
\email{dchgomes@sas.upenn.edu}
\affil{Department of Physics \& Astronomy, University of Pennsylvania, 
209 S.\ 33rd St., Philadelphia, PA 19104}

\author[0000-0002-8613-8259]{Gary M. Bernstein}
\affil{Department of Physics \& Astronomy, University of Pennsylvania, 
209 S.\ 33rd St., Philadelphia, PA 19104}

  \begin{abstract}
We explore the potential of an array of $\mathcal{O}(100)$ small fixed telescopes, aligned along a meridian and automated to measure millions of occultations of Gaia stars by minor planets, to constrain gravitational signatures from a ``Planet X"  mass in the outer solar system.  The accuracy of center-of-mass tracking of occulters is limited by  photon noise, uncertainties in asteroid shapes, and Gaia's astrometry of the occulted stars. Using both parametric calculations and survey simulations, we assess the total information obtainable from occultation measurements of main-belt asteroids (MBAs), Jovian Trojans and trans-Neptunian objects (TNOs). We find that MBAs are the optimal target population due to their higher occultation rates and abundance of objects above LSST detection thresholds. A 10-year survey of occultations by  MBAs and Trojans using an array of 200 40~cm telescopes at 5~km separation would achieve $5\sigma$ sensitivity to the gravitational tidal field of a $5M_\oplus$ Planet X at 800~AU for $>90\%$ of potential sky locations. This configuration corresponds to an initial cost of $\approx\$15$ million. 
While the survey's sensitivity to tidal forces improves rapidly with increasing number of telescopes, sensitivity to a Planet X becomes limited by degeneracy
with the uncertain masses of large moonless TNOs. The 200-telescope survey would additionally detect $\approx1800$ TNO occultations, providing detailed shape, size, and albedo information. It would also measure the Yarkovsky effect on many individual MBAs, measure masses of many asteroids involved in mutual gravitational deflections, and enable better searches for primordial black holes and departures from General Relativity.
\end{abstract}
\section{Introduction}
Stellar occultations, \ie, transits of an occulter object that temporarily blocks the light from a background star, provide a convenient method for studying physical properties and orbital dynamics of solar system objects. Historically, potential uses of the method have long been discussed, as seen in Leonhard Euler's application for determining longitude with stellar occultations by the Moon \citep{Euler1749} and Rev. John Michell's mention of measuring stellar angular diameters using occultations by Venus \citep{1767RSPT...57..234M}. The 20th century witnessed an increased interest in the study of planetary atmospheres via occultations, building upon theoretical development by \citet{1903AN....164....5P} and \citet{1929JO.....12....1F}. Towards the end of the century, successful observations allowed the discovery of Uranus’ rings \citep{ELLIOT1977} and the detection of Pluto’s atmosphere \citep{Hubbard1988}.

The advent of high-precision stellar catalogs has expanded possibilities in the field, especially in the study of small solar system bodies. These measurements allow determination of asteroid shapes, which can potentially constrain solar system formation models, and provide better orbital tracking than attainable through direct observation of the target. Recent effort in understanding the outer regions of the solar system has motivated the prediction and observation of trans-Neptunian object (TNO) occultations through projects such as the RECON network \citep{recon}.  

Targeted occultation surveys rely on previous knowledge of each target's orbit with precision comparable, at least, to the size of the telescope array. \citet{Ortiz2019} note that, though Gaia astrometry is at $\mu$as level, many TNO orbits are still uncertain to 300~mas. At a distance of 40~AU, this corresponds to $\sim 9000$~km shadow path uncertainties, which implies a low probability of successful event prediction. Some objects, however, have been better constrained—---New Horizons extended mission target Arrokoth was successfully observed in three occultation campaigns \citep{Buie_2020} following orbital determination from HST astrometry, and other TNOs closely observed by the New Horizons spacecraft also had their orbital accuracy improved \citep{Porter_2022}. As the \textit{Legacy Survey of Space and Time (LSST)} commences and proceeds over its c.\ 2025--2035 period, nearly all of the minor bodies it detects will accrue sufficiently accurate orbital data to predict occultation tracks to $O(100)$~km, hugely increasing the efficiency of targeted occultation measurements.

Main-belt asteroids (MBAs) and Jovian Trojans are more accessible targets, since they are closer to us and have larger apparent paths in the sky. These populations are also of interest in outer solar system studies, particularly due to their potential use as test particles in the search for distant gravitational signals, which could provide constraints on different ``unseen planet'' hypotheses (generally referred to as Planet X), such as the Planet Nine hypothesis evoked to explain a perceived orbital clustering of TNOs ~\citep{Trujillo_2014,Sheppard_2016,Batygin_2016,Brown_2016}. \citet{Rice_2019} suggested a network of $\sim$ 2000 telescopes spread over continental United States measuring occultations of Jovian Trojans, with the goal of detecting perturbations by a putative Planet X. Inspired by this idea, in a previous study \citep{Gomes_2023}, we explored the ability of current and planned LSST astrometric data to constrain the presence of a distant tidal field, finding that LSST astrometry of Jovian Trojans does not significantly improve  Planet X mass uncertainties. The possible contribution of an occultation array, however, remained unexplored beyond Rice and Laughlin's initial foray.

In the current paper, we propose a fixed array of telescopes aligned roughly along a meridian, automated to detect targeted occultations of Gaia stars by asteroids from any of the three main populations---MBAs, Jovian Trojans and TNOs. Starting with LSST-level astrometry for initial orbit determination, the array would narrow down center-of-mass positions through multiple occultation events per asteroid over a period of $\approx 10$~yrs. We explore which array design and target choices would yield better constraints on the tidal fields of a distant point mass.

This paper will focus on optimizing a telescope array for minor-body astrometry and tidal-field detection, but an array of $\gtrsim100$ agile telescopes could also advance many other science goals.  The astrometric data would also improve the solar system ephemeris generally, constrain the flux of asteroid-mass primordial black holes \citep{Tung2024}, and vastly improve knowledge of the sizes of asteroids and their non-gravitational forces that drive much migration.  A review of knowledge gained on TNOs through occultations is given by \citet{Sicardy24}, and in Section~\ref{sec:results} we show that such an array would measure the shapes and albedos of $\approx 100\times$ more TNOs than have been characterized to date.
Individual stations would be available for other observations when they are not active for an occultation, enabling high-duty-cycle monitoring for asteroid impacts, supernova shock breakouts, and many other transient or variable phenomena occurring on time scales or cadences inaccessible to LSST.

\section{Hardware characteristics and costs}
We will assume that observations are made by an array of $N\gtrsim 100$ telescopes, each with diameter $d\sim0.5$~m, separated by distance $D\sim1$~km, spanning a length of $L=ND\gtrsim100$~km (symbols and relevant default values are listed in Table~\ref{tab:symbol}).  The array should be oriented roughly north-south, perpendicular to the typical (ecliptic) apparent motion of the occulters, but it is not necessary to have precisely spaced or oriented stations in the array.  
We will assume throughout that occultations are measured through a filter centered at $\lambda_0=600$~nm, spanning $400\,{\rm nm} < \lambda < 800\,{\rm nm},$ which is a common choice for maximizing signal-to-noise ratios ($S/N$) in the presence of atmospheric and zodiacal backgrounds.  We will further assume a nominal survey duration of $T=10$~yr.

Other important characteristics of the telescopes include the overall quantum efficiency $\eta$ of the atmosphere/telescope/aperture/detector combination, for which we assume 50\%.  The latitude, maximum observable airmass $X_{\rm max},$ and duty cycle $f_{\rm duty}$ are important to determining the yield of the occultations, where the latter is the fraction of occultations that pass over (or under) the array that can actually be observed.  We will default these to $\pm30\arcdeg,$ 2.3, and 0.21, respectively.  For an observatory at latitude $\pm30\arcdeg,$ $f_{\rm duty}$ is limited to $<0.32$ by the fraction of near-ecliptic sky that is visible at airmass $<X_{\rm max}=2.3$ We reduce this by a factor $0.65$ to allow for losses due to clouds, down time, and bright twilight or moonlight that might be present---our simulations described in Section~\ref{sec:sims} will explicitly check each potential event against solar, lunar, and weather conditions rather than assuming an $f_{\rm duty}$ value.

The detectors on each telescope will need to read out the flux of the occulted star at $\gtrsim200$~Hz in order to resolve the $\approx10$~ms transition from full to zero shadow in egress/ingress.  Read noise below $1e^-$ will be needed.  A potential solution is a single-channel photon-counting device, such as an avalanche photodiode.  This detector would be paired with a CCD or CMOS detector with field of view of several arcminutes, sufficient to find several stars in the Gaia catalogs for any pointing in the sky.  This ``finder'' detector would be read out, the image analysed to determine the current telescope pointing, and then the pointing adjusted to place the target star onto the photon counter's aperture.  Thus the pointing system needs to be capable of sub-arcsecond accuracy only for differential movements of a few arcminutes.  The telescope mount requires only several-arcminute accuracy for longer slews.

The occultation events will last only $\lesssim 1$~second for MBAs and Trojans, and up to $\approx 1$~minute for TNOs.  We thus envision a very rapid automated cadence, in which each station can independently slew to a target, expose and analyze a registration image, shift the target onto the photon counter, and record several seconds of data, then move on to the next target in a total of $\lesssim 2$~minutes.  It is possible that subarray readout of a CMOS ``finder'' detector could have low enough noise to obviate the need for a photon counter.

To minimize costs, each station should be as autonomous, reliable, and inexpensive as possible.  We envision each station being powered by a solar array and battery, and transmitting at least station-keeping information by cellular links, so no utility hookups are needed.  If cellular transmission of the observing data is too expensive,  these data could be stored on local physical media that are collected periodically by maintenance crews.  A modular telescope/mount/enclosure unit should be small enough that several can be placed on a truck, so that a crew travelling along the array can swap out units and bring them back to a central lab for repairs and maintenance.  Similarly for computing and power systems.

\subsection{Cost equation}
We will make very crude estimates of array capital costs as
\begin{equation}
  \text{cost} = N\left[ C_s + C_t\left(\frac{d}{0.5\,\text{m}}\right)^2\right],
  \label{eq:cost}
\end{equation}
where $C_s$ is a fixed cost per station for site preparation, enclosure, detectors, power, computing, and communications; and $C_t$ is a telescope cost, which we assume will scale with the collecting area.  Some rough guesses are:
\begin{itemize}
\item {\bf Telescope:} Ritchey-Chretien telescopes from Planewave Instruments\footnote{\url{https://planewave.com/collections/all-telescopes/}} have costs that scale roughly linearly with collecting area for $d\lesssim0.7$~m, with a telescope$+$mount cost of $\approx\$60,000$ for $d=0.5$~m.  We thus adopt a default of $C_t=\$60$k.  Ordering at $>100$-unit scale might make this and other components substantially cheaper.  Catadioptric telescopes are less expensive and available for $d\lesssim0.4$~m.
  
\item {\bf Detectors:} A CMOS or CCD ``finder''  with $\approx10\arcmin$ FOV would not require particularly large format or high performance, so commercial off-the-shelf hardware would suffice.  Commercial single-channel photon-counting devices should work well too; we want to push for high QE, but the timing requirements (millisecond levels) and count rates (100--1000 Hz) that we require are more relaxed than typical specifications.  The two detectors might cost $\approx\$15$k per station total.

\item An {\bf enclosure} would need to be designed and replicated; a simple
  hinged-top box would be sufficient, and would not need to be large
  enough to contain a human---the sides could open for maintenance.
  Perhaps this could be done for \$10k per station.

\item \textbf{Computing, solar power, and cellular data link} would cost
  a few \$k per station.

\item \textbf{Site prep} would comprise a concrete pad and a security fence and cameras, perhaps \$10k.

 \item Least predictable would be the costs of negotiating and acquiring the leases for the sites.  We ignore this important element for now, because any believable estimate would depend heavily on choice of site and require research well outside of our astronomical expertise.
 \end{itemize}
In sum, a crude guess is hence $C_s=\$40$k and $C_t=\$60$K, leading to a \$10M cost for a 100-station array.
We have neglected design costs (including software), which probably are independent of the array specifications. The above costs are all for capital outlays.  There would of course be ongoing operating costs for maintenance/personnel.  At most observatories these are 5--10\% of capital cost annually, accumulating to be similar to capital costs over the project lifetime.  However, this unusual observatory might be more expensive because of its geographically distributed nature, or less expensive because its replication of components makes maintainance more efficient.  An engineering and cost study is needed.

\section{Parametric analyses}
In this Section we describe a simplified parametric estimation of the cost and performance of an occultation array, which we then use to study the constraints and optimization on array parameters to produce useful information on gravitational perturbations subject to cost bounds.
In Section~\ref{sec:results} we will re-evaluate the optimization and performance using more thorough simulations of survey scenarios, guided by the results of these simpler parametric analyses.

The forecast of constraints on a given gravitational signature requires answers to the following questions:
\begin{enumerate}
\item What is the uncertainty in position $\sigma_x$ of the target minor body attained from a single occultation event?
\item How many events (at what uncertainties) would the array be able to attain on a typical object?
\item How many test particles (occulters) can be tracked from the selected small-body population?
\item What constraints does the suite of observations place upon the gravitational anomaly?
\end{enumerate}
For a given observing scenario and target population, a useful figure of merit is the total positional information $\sigma_{\rm tot}$ defined by
\begin{equation}
  \label{eq:sigtot}
  \sigma_{\rm tot}^{-2} = \sum_{i\in\text{targets}} \sum_{j \in \text{occulations of}\,i} \frac{1}{\sigma^2_{x,ij}},
  \end{equation}
  with $\sigma^2_{x,ij}$ being the measurement uncertainty on the position of test particle $i$ during its $j$th occultation event.  The resulting $\sigma_{\rm tot}$ can be compared to the few-meter scale of information from ranging to spacecraft orbiting Jupiter and Saturn, or few-cm scale information on the range to Mars, keeping in mind that the displacements caused by an anomalous tidal force on a test body will scale roughly as $\Delta x \propto a_T^4,$ where $a_T$ is the target's semi-major axis.
  
\subsection{Test particle populations}
\begin{figure}
  \centering
  \includegraphics[width=0.8\textwidth]{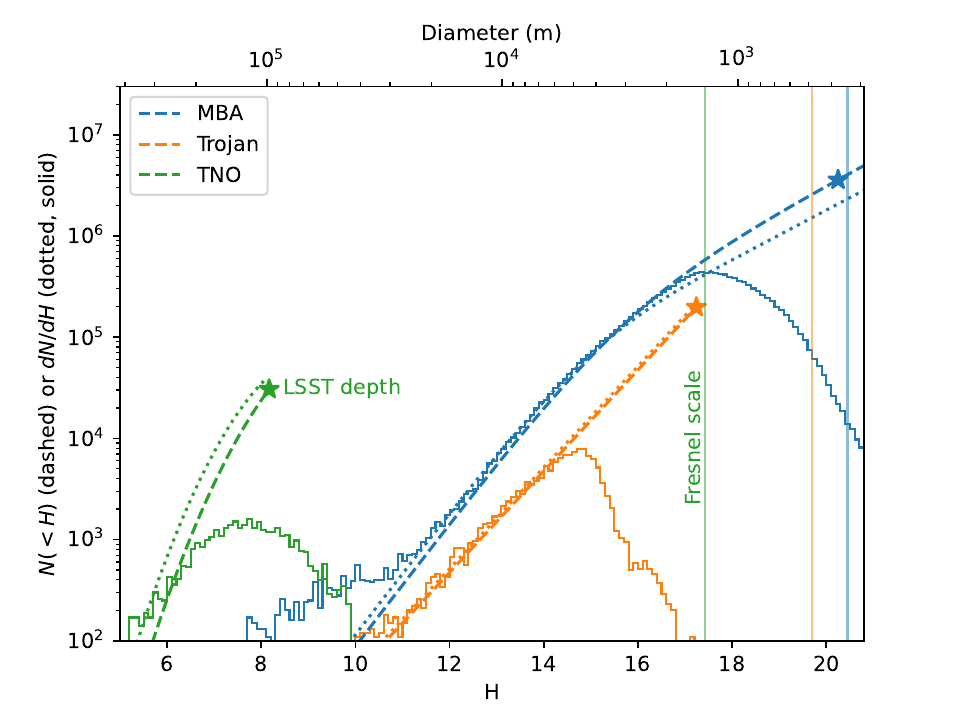}
  \caption{The dashed and dotted lines show the assumed cumulative and differential counts, respectively, for the MBA, Trojan, and TNO populations.  The histograms show the differential counts $dN/dm$ of objects currently listed by the Minor Planet Center.  Each cumulative curve terminates at a star marking an apparent magnitude $m_V=24.4,$ which is roughly the limit at which objects can be detected in single visits by LSST.  The vertical lines mark the Fresnel scale for each population.  Note that our models are not intended to match the excesses of larger bodies.}
  \label{fig:counts}
\end{figure}

We examine as target populations each of the three most populous reservoirs in the Solar System: the main-belt asteroids (MBAs), the Jovian Trojan asteroids, and the trans-Neptunian objects (TNOs).  In this section we consider members of each population to share a common semi-major axis $a_T=2.6, 5.2,$ and $42$~AU for the MBAs, Trojans, and TNOs, respectively.  Aside from $a_T,$ the other critical characteristic of the target population is $N_T(>d_T),$ the number of members larger than diameter $d_T.$  Figure~\ref{fig:counts} plots the values we have assumed for each population.  Simple power laws are good fits to the regime where the MBAs and Trojans cataloged by the MPC\footnote{\url{https://www.minorplanetcenter.net/data}} are largely complete. For fainter MBAs, we adopt the formula used in the initial LSST predictions \citep{LSST_science_book}, and for Trojans we extrapolate the power law fit. For TNOs, we adopt the rolling power law for $dN/dH$ found by \citet{Pedro24} 
to fit the $6<H_r<8.2$ range.  We forecast that the known populations in future years will be dominated by objects detected by LSST.  The expected $5\sigma$ detection levels for single-epoch LSST exposures in nominal conditions (dark, zenith, median seeing)\footnote{\url{https://pstn-054.lsst.io/}} are equivalent to $m_V=24.4$~mag, which we hence consider a bound on the usable occulter population.
There are $> 10$ times more usable MBAs than Trojans, and potentially $10^2$ more MBAs than usable TNOs---but the physical displacements induced on a target by a tidal force scale as $a_T^4$ (for fixed number of orbits), so it is not \textit{a priori} obvious which population will be most informative.

\subsection{Measurement errors}
\label{meas_err_1}
There are three important sources of uncertainty in the transverse position of the center of mass of a given target during an occultation, which we will call \emph{shape} noise ($\sigma_{\rm SN}$), \emph{photon} noise ($\sigma_\gamma$), and \emph{Gaia} uncertainties $\sigma_G.$ 

The ultimate limit to the determination of orbits through occultations is the shape noise, our ignorance of the target's projected center of mass location relative to the measured chords.  Even perfect knowledge of the silhouette of the target would yield an uncertain center-of-mass (CM) estimate because of ignorance of the depth dimension, and variations in internal density cause additional CM shifts relative to the geometric centroid.  
In Section~\ref{sec:shapenoise} we justify an estimate that $\sigma_{\rm SN} = \alpha_S d_T,$ with a nominal $\alpha_S=0.05$ giving the RMS ratio of the along-track CM distance from the center of a single chord to the mean radius of the target. The cross-track uncertainty from a single chord CM is larger, with $\sigma\approx{\rm min}(d_T,D)/\sqrt{12},$ since a simple estimator is that the CM has a uniform probability of being somewhere in a region of width ${\rm min}(d_T,D)$ centered on the observed chord.  In this Section we will examine only the along-track information; the full simulation will use cross-track and multi-chord information as well.

For $\sigma_G$, we use the estimated accuracy of Gaia DR5 astrometric solutions provided by ESA.\footnote{\url{https://www.cosmos.esa.int/web/gaia/science-performance}}

Appendix~\ref{sec:diffraction} derives the uncertainty $\sigma_\gamma$ in the center of a chord due to Poisson noise in the arrival of photons from both the occulted star (having unocculted photodetection rate $n_\star$) and some foreground/background count rate $n_b$ due to atmospheric and zodiacal foreground signals, as well as sunlight reflected from the occulting asteroid. The background ratio is defined as $B = n_b/n_\star$.  For $B\rightarrow0,$ bright stars, and large occulters, this uncertainty becomes $\sigma_\gamma = F / \sqrt{2n_\star t_F},$ where  $F=\sqrt{a_T\lambda_0/2}$  is the \emph{Fresnel scale} which characterizes the distance over which the occulter's shadow  transitions from fully bright to fully dark.  This transition region takes a time $t_F=F/v_\perp$ to sweep over the telescope, where $v_\perp$ is the relative speed of the occulter and the observer transverse to the line of sight.  The Fresnel scales $F$ and $t_F$ are $\approx 340$~m and 11~ms, respectively, for typical MBAs, growing as $\sqrt{a_T}$ for more distant occulter samples.

This best-case result is degraded if:
\begin{enumerate}
\item   The occulter has $d_T$ below the Fresnel scale, which leads to shallow occultations that are harder to measure, tending toward $\sigma_\gamma \propto d_T^{-5/2}.$  This essentially limits the available gravitational information to objects with $d_T>F=340 (a/2.6\,\text{AU})^{1/2}$~m.
\item The star is too faint to deliver at least several photons during $t_F,$ or the background is $B\gtrsim0.1$ and similarly degrades the $S/N$ during the ingress/egress transitions.
\item The occulter becomes large and bright enough that its reflected light fills in the occultation light curve significantly.
\end{enumerate}
\eqq{eq:sigmagamma} gives an approximation to $\sigma_\gamma$ that spans the different regimes of source brightness, background level, and occulter size.  An important consideration is that the occultation must be long and deep enough to be detected in the first place.  We will require that $d_T>4 \sigma_\gamma$ to approximate this condition.

\begin{figure}
  \centering
  \includegraphics[width=0.49\textwidth]{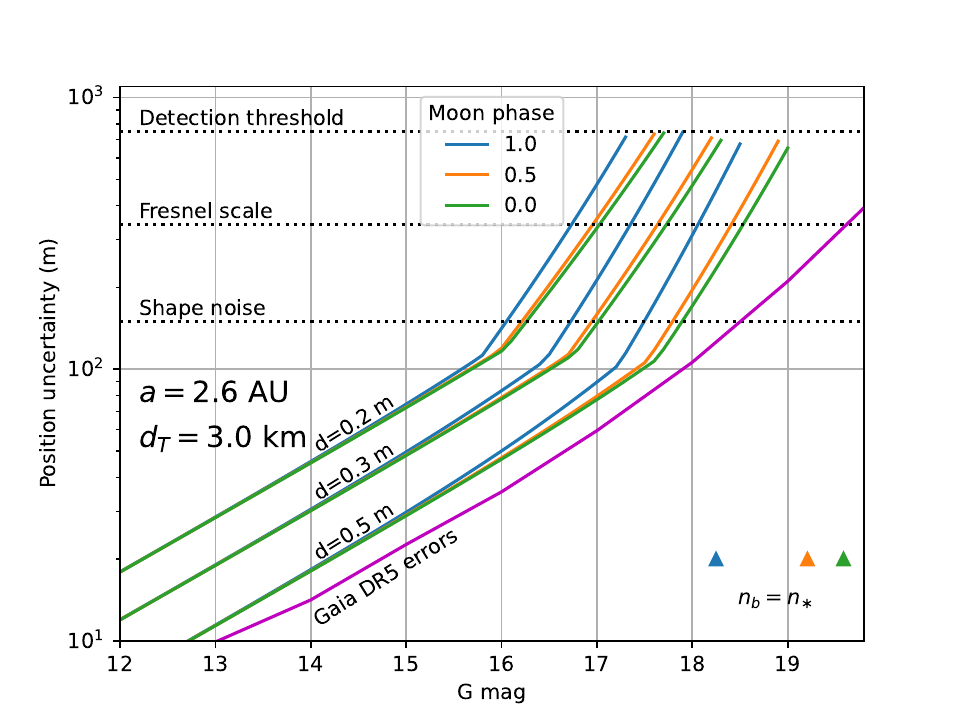}
  \includegraphics[width=0.49\textwidth]{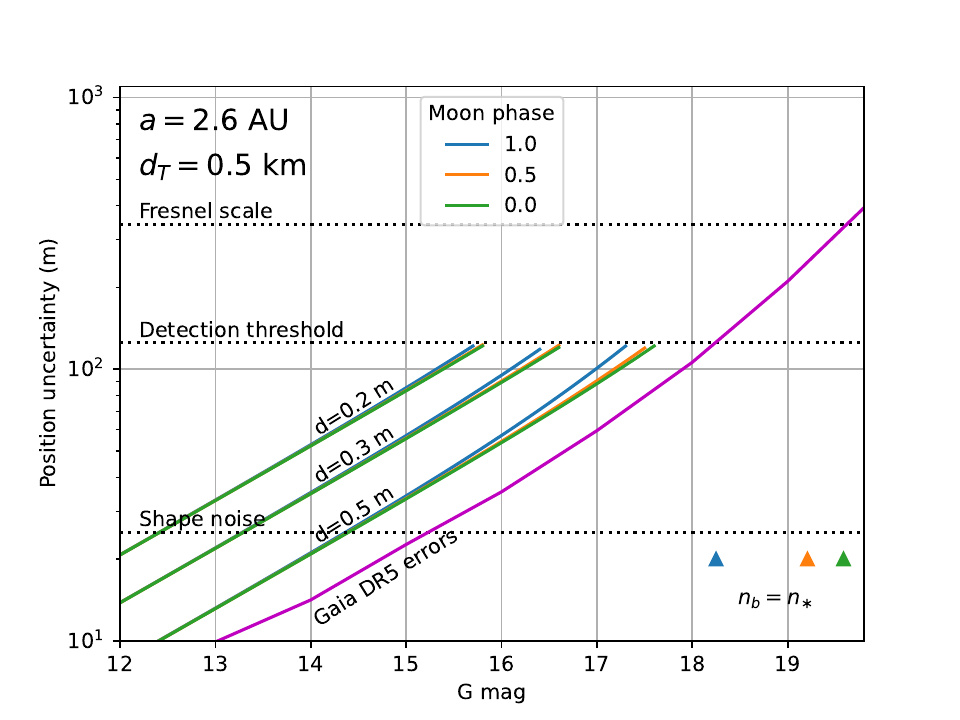}
  \caption{Sources of uncertainty in the along-track asteroid position from a single occultation chord are plotted vs magnitude of the occulted star. At left is for an occulter diameter $d_T=3$~km, at right $d_T=0.5$~km, both sources at distance 2.6~AU.  Each set of lines shows values of the photon noise $\sigma_\gamma$ for new, quarter, and full Moons, and three different telescope diameters $d$ are plotted in each case.  The magenta curve plots the expected final Gaia positional uncertainties $\sigma_G$, and the horizontal line marked ``shape noise'' marks the floor $\sigma_{\rm SN}$ set by the unknown shape and density distribution of the occulter.  The ``detection threshold'' marks the uncertainty $\sigma_\gamma=d_T/4$ that we consider an upper limit for detectable occultations, and the $\sigma_\gamma$ curves terminate at this level.
    The Fresnel scale is also plotted, and the small triangles mark the $G$ magnitudes at which background counts equal star counts for different moon phases.  For the 0.5~km occulter, near the diffraction scale, almost all occultations are limited by photon noise, and the sky brightness is unimportant.  For the larger occulter, fainter Gaia stars yield detectable occultations, and the shape noise limits the measurement accuracy for much of the detectable regime.}
  \label{fig:sigx}
\end{figure}

Figure~\ref{fig:sigx} plots the resultant expected uncertainties $\sigma_\gamma, \sigma_G,$ and $\sigma_{\rm SN}$ vs source Gaia magnitude $m_G$ for MBAs of a few different sizes, to illustrate the domains in which each is important.  One  conclusion is that telescopes should have $d_T\lesssim0.5$~m; larger telescopes would lower $\sigma_\gamma,$ but then $\sigma_G$ would dominate the total $\sigma_x$ quadrature sum, negating the gain of the larger, more expensive telescopes.

\subsection{Event rates}
An estimate of the number of observed occultations of stars with $ m_G < m_{G,\text{max}}$ per target asteroid during the survey is obtained by finding the sky area subtended by the outline of the asteroid's track across the sky, multiplying by
the sky density of potential stellar backlights, and by the duty cycle of occultation observations:
\begin{align}
  \langle N_e \rangle & = f_{\rm duty}\times  n_G(<m_{G,\text{max}})\times T \left\langle\text{apparent motion} \right\rangle 
                   \left( \frac{\text{effective array length}}{a_T}\right) \\
                    & \approx 10 \left(\frac{T}{10\,{\text yr}}\right) \left(\frac{f_{\rm duty}}{0.21}\right) 10^{0.3(m_{G,\text{max}}-18)} \left(\frac{\sin^{-1}(1\,{\rm AU}/a_T) / a_T}{\sin^{-1}(1/2.6) / 2.6}\right)
                      \left(\frac{N\times\text{min}(d_T,D)}{200\,\text{km}}\right).
                      \label{eq:eventrate}
\end{align}
The final expression makes use of the following estimates: first, we estimate the mean sky density of useful Gaia stars within $\pm10\arcdeg$ of the ecliptic as $n_G(<m_G)\approx 3500\times10^{0.3(m_G-18)}\,\text{deg}^{-2}.$ This estimate does not count as ``useful'' those Gaia stars in crowded low-Galactic-latitude regions.

Second, we approximate the average annual length of a target's apparent sky track, constrained to those portions with solar elongation $\beta > \beta_{\rm min}=50\arcdeg,$ as $2 \sin (1\,\text{AU}/a_T),$ which is accurate to $<10\%$ for circular orbits.

Not all positional information on a target can be used to constrain unexpected gravitational accelerations.  There are six degrees of freedom in each target's initial state vector, which must also be solved using the occultation data.  Previous data (nominally from LSST) would be available too, but will probably be $\gtrsim10\times$ less accurate than the occultation data and hence not useful at occultation-survey accuracy---otherwise we would not be doing an occultation survey!  We therefore assume that, for a given asteroid, the three most precise occultation measurements will be used to constrain the initial state, and information on perturbations will begin to accrue only from the 4th most accurate occultation observation.  Thus a target which does not produce $\ge4$ measurable occultations does not contribute at all to the $\sigma_{\rm tot}$ figure of merit.  When forecasting the actual number of occultations, we assume each target draws from a Poisson distribution with mean given by \eqq{eq:eventrate}, rather than assigning every asteroid the mean number of observed occultations.

\eqq{eq:eventrate} shows that $\langle N_e \rangle \propto a_T^{-2},$ thus the Trojans and particularly the TNOs will be starved for occultations, with $\langle N_e \rangle <4$  unless we increase the array length well beyond 200~km.  
It is also clear that the number of observed events $N_e$ scales with the number of telescopes $N$ at a fixed choice of spacing, so our goal quantity will scale as $\sigma_{\rm tot} \propto N^{-1/2},$ and in fact will scale more strongly than this because of the threshold $N_e\ge 4$ required to make a test particle useful.  

\subsection{Results: MBAs}
\label{sec:simpleMBA}
\begin{figure}
  \centering
  \includegraphics[width=0.49\textwidth]{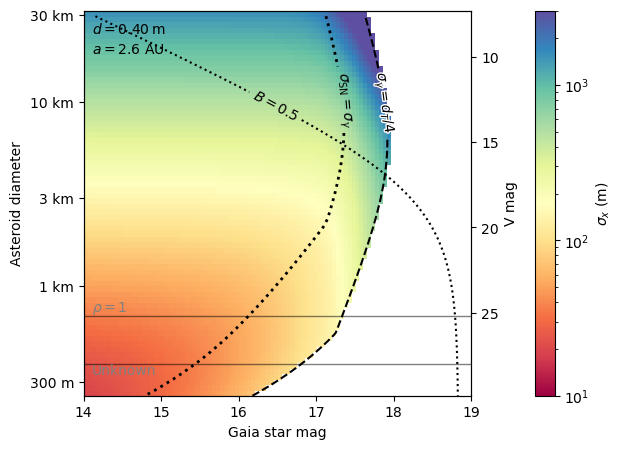}
  \includegraphics[width=0.49\textwidth]{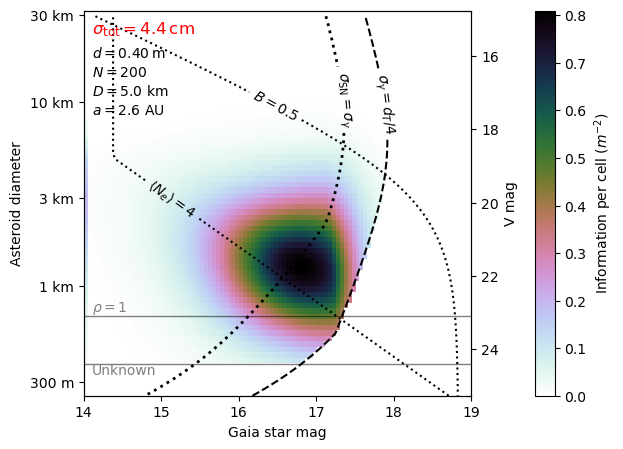}
  \caption{These figures illustrate the performance predicted by the parametric model for an occultation array targeting MBAs using a configuration of $N=200$ telescopes of diameter $d=0.4$~m at $D=2$~km spacing, near the optimum for a \$15M nominal capital cost.  The uncertainty on the RMS size of an anomalous gravitational displacement after a 10-year survey is just $\sigma_{\rm tot}=4.4$~cm.
At left is the total measurement uncertainty in the occulter position for single-chord observations, vs.\ the occulter diameter and magnitude of the occulted star.  Overlaid are several important dividing lines: asteroids below the $d_t=2F$ line have radius below the Fresnel scale, only partially blocking stellar light.  Those below the ``Unknown'' line are too faint to be found in the nominal LSST single-epoch data.  Those to the right of the $B=0.5$ contour are substantially degraded by ``background'' light from the night sky and the asteroid's reflected light.  Those to the left of the $\sigma_\gamma=\sigma_{\rm SN}$ contour are limited by shape noise in determining the center of mass of the asteroid.  Those to the right of the $\sigma_\gamma=d_T/4$ curve have too much photon noise for the occultation to be securely detected.  The right-hand plot shows the density of information on gravitational perturbations in the $d_T-m_G$ plane, which is heavily concentrated in the region with the most numerous occultations---the asteroids near the Fresnel scale, with the faintest stars that produce detectable occultations.  The $\langle N_e \rangle=4$ contour marks the magnitude of the star that will on average be the 4th-brightest occulted by an asteroid of a given size.  Occulters with $N_e<4$ produce negligible information on gravitational perturbations.}
\label{fig:simpleMBA}
\end{figure}

Figures~\ref{fig:simpleMBA} show, for one scenario of an MBA survey, the results of the above assumptions on a grid of MBA diameter vs Gaia source magnitude.  The left panel shows the expected $\sigma_x$ per occultation for the $(d_T,m_G)$ combination.  The right panel weights the $\sigma_x^{-2}$ information per occultation in each region in $d_T$-$m_G$ space by the number of asteroids and the number of usable occultations per asteroid, to give the total information density in $d_T$-$m_G$ space, following \eqq{eq:sigtot}.  An occultation is assumed to be useful only if $\ge3$ observations with lower $\sigma_x$ have been accrued for that object.

The projected value of $\sigma_{\rm tot}$ in this scenario is just 4.4~cm, similar to the accuracy available from Mars ranging, with a nominal \$15M capital cost.  This is quite encouraging, since the tidal field of a $5M_\oplus$ Planet X at 800~AU will cause a shift of $\approx5e_T$~meters from apsidal precession over 10 years (where $e_T$ is the occulter's eccentricity) and a quadrupolar position oscillation with amplitude $\approx\pm10$~cm on each orbit.

These plots show how the region of useful information is bounded at the faint-$m_G$ end by having sufficient statistics to detect the occultation, ($\sigma_\gamma<d_T/4$). This bound moves to fainter $m_G$ with larger telescope diameters $d,$ and is independent of the number of stations $N$. The high-information region is bounded at the bright-$m_G$ end by the need to accrue enough occultations over 10 years to determine the state vector and begin to extract information on perturbations ($N_e>4$). This bound improves linearly with $N$ and with the survey duration $T$,  and is largely independent of $d.$   As the target diameter $d_T$ decreases, the number of available occulters and available information increase rapidly, until these two constraints converge, which by coincidence is near the Fresnel scale in this scenario.  The majority of information is arriving from asteroids near the $\langle N_e\rangle=4$ line, with some targets below the line still providing their contribution if they end up with $N_e\ge4$ occultations because of Poisson fluctuations in the number of occultations over the survey.
The bulk of the information on gravitational perturbations comes from MBAs slightly above the Fresnel scale, since the number of such bodies increases rapidly with smaller diameter, and likewise is concentrated in the faintest source-star magnitudes whose occultations are detectable.

The parametric treatment of this section allows rapid exploration of the space of survey characteristics.  We observe that configurations optimized under a cost constraint have the following characteristics:
\begin{itemize}
\item The tradeoff between telescope diameter $d_T$ and number of telescopes $N$ at fixed cost minimizes $\sigma_{\rm tot}$ when the cost of the telescope $C_t(d_T/0.5\,{\rm m})^2$ is 10--20\% higher than the fixed station costs $C_s$.  In other words, onde should spread the funds over as many (smaller) telescopes as possible until the per-station costs are about to dominate.  When $C_s=\$40{\rm k}$, this leads to optimal telescope size $d_T\approx0.45$~m.
\item The constraining power of the array grows rapidly with the total cost cap, $\sigma^{\rm tot} \propto \text{cost}^{-1.3}.$ Higher capital funds allow larger $N$, which benefits the performance in two ways: first, the total number of observable occultations increases linearly with $N$.  Second, the 4-brightest occultation moves to brighter stars, which means that lower-$\sigma_x$ measurements can now be applied to gravitational constraints.

\item Increasing the survey duration $T$ increases the constraining power at an even higher exponent, since increasing $T$ and increasing $N$ reduce the $\sigma_{\rm tot}$ at the same rate---but increasing $T$ gives gravitational anomalies more time to act, raising their resultant displacements.

\item Reducing the fixed costs per station is highly beneficial, roughly as $\sigma_{\rm tot} \propto C_s^{-0.5},$ at fixed total costs since this moves the optimization toward more, smaller telescopes, improving $\langle N_e \rangle$.  

\item The right panel of Figure~\ref{fig:simpleMBA} shows that LSST is capable of discovering nearly all of the MBAs that would make useful test particles.

\item The Figure also shows that most of the information is coming from occultations that are \emph{not} strongly affected by background light.  This is good news in that it makes the survey less sensitive to the degraded seeing and night-sky brightness that a geographically dispersed array might encounter.

\item It is also true that much of the MBA information is from objects close to the Fresnel size and ``lucky'' to obtain 4 measured transits.  Both of these place stress on our simplified assumptions about the amount of information extractable from each occulter, and the latter might reduce our robustness against other nuisance effects such as non-gravitational forces.  The simulations in Section~\ref{sec:sims} will test the robustness of the extracting of gravitational inferences from the positional data.
\end{itemize}

\subsection{Results: Trojans}
\label{sec:simpleTrojan}
\begin{figure}
  \centering
  \includegraphics[width=0.49\textwidth]{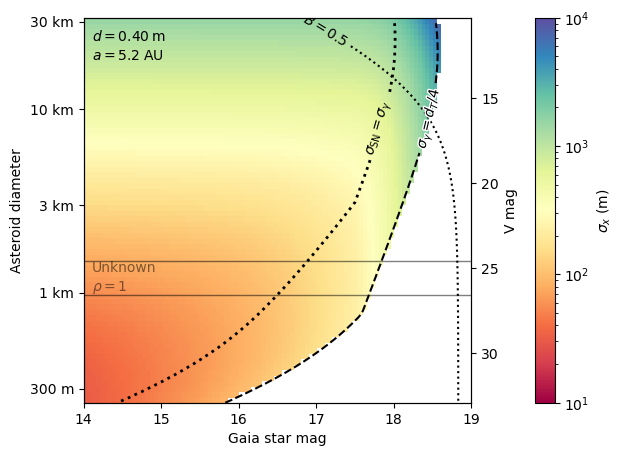}
  \includegraphics[width=0.49\textwidth]{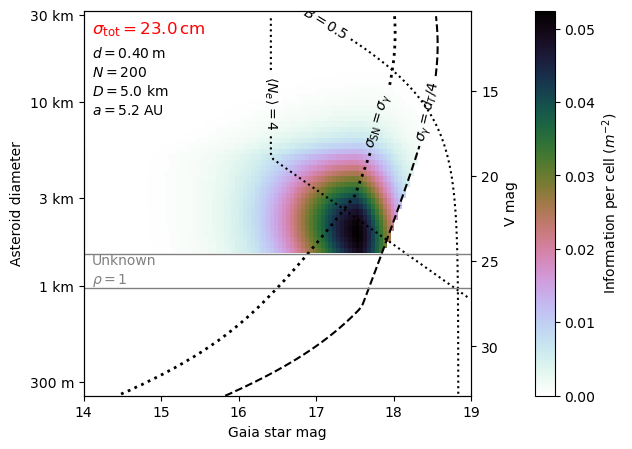}
  \caption{Summaries of the per-occultation measurement errors (left) and total information (right) for an occultation survey targeting Jovian Trojans.  All curves have the same meaning as for the MBA case in Figure~\ref{fig:simpleMBA}, except that the color scales are changed.  The main configuration change is to move the telescopes further apart from each other to increase the occultation rates for the larger-diameter, sparser Trojan population.  Analysis is in Section~\ref{sec:simpleTrojan}.}
\label{fig:simpleTrojan}
\end{figure}

Using the Jovian Trojans as test bodies instead of MBAs changes the optimization of an occultation array in important ways, even though the $2\times$ increase in Earth-occulter distance changes the Fresnel scale by only $\sqrt{2}.$  One issue is that, per Figure~\ref{fig:counts}, there are $\approx5\times$ fewer Trojans of a given size than MBAs.  Furthermore, a Trojan of a given size is 3~mag fainter than its MBA counterpart, which means that LSST's nominal survey is \emph{not} capable of finding all Trojans above the Fresnel scale, as can be seen in Figures~\ref{fig:simpleTrojan}.  Most importantly, the sky area subtended by the annual track of a Trojan is $4\times$ lower than for a similar-sized MBA, which means we must re-optimize the array to obtain $\langle N_e\rangle\ge4$ for a substantial fraction of the observable Trojans.

We therefore investigate changing the telescope spacing from $D=2$~km for the MBAs to 5~km for the Trojans, which will increase the number of observable occultations for Trojans with $d_T>2$~km---which is, in fact, all of the LSST-discoverable sources.  Figures~\ref{fig:simpleTrojan} illustrate the result of a configuration near the new optimum for a \$15M nominal cost.  The number and size of telescopes resemble the MBA optimum, but spread out over 1000~km.  This wider-spaced array might be harder to maintain; it would, however, also be able to conduct the MBA survey, with the caveat that many forecasted MBA occultations would be ``duds'' that pass between stations, leading to a higher workload for the array.

The scaling relations of the optimal $d_T$ and of $\sigma_{\rm tot}$ vs. the cost, duration, and per-station costs remain essentially the same as the MBA case.  Most of the occultations still occur in low-background situations. Some practical differences are, however, that
\begin{itemize}
\item Most of the information is now arising from occultations that are photon-noise limited rather than shape-noise limited.
\item The usable occulters are above the Fresnel size, so occultations will be deeper and will last longer.
\item The total number of Trojan occultations would be many times lower than the number of MBA occultations.
\end{itemize}

The reduced number of Trojan targets and the need to use larger occulters both act to reduce the statistical power of the Trojan survey relative to the MBA.  Our \$15M scenario forecasts $\sigma_{\rm tot}=23$~cm, which is $\approx5\times$ less sensitive to displacements than the MBA survey.  It is also true, however, that the apsidal precession from a Planet X tidal force over a fixed time T scales as $a_T^{5/2}$ and the quadrupolar oscillation amplitude scales at $a_T^4,$ factors of 5.6 and 16 larger for Trojans than MBAs, respectively.  Hence our parametric investigations suggest similar constraining powers for studies of the two populations---and considerable gain if we can do most of both surveys with one array.

\subsection{Results: TNOs}
\label{sec:simpleTNO}
An occultation survey of TNOs for the purposes of constraining tidal forces would face more extreme versions of the challenges of a Trojan survey: first, the number of targets discovered and with good orbit determinations from LSST is very low, with a recent estimate of 30,000 (J. Kurlander, private communication), with the smallest of them near $d_T=100$~km.  Thus the shape-noise limit alone will degrade per-event accuracy to well above the Fresnel scale.  The good news is that the longer, deeper occultations will allow detection of occultations of fainter Gaia stars.  This increase in event rate will, however, be overwhelmed by the huge decrease in the area subtended by the apparent motion of the occulters.  Anticipating that we will need maximal array cross-section to obtain $N_e\ge4$ occultations on the scarce targets, we consider a global-scale array with $100$ telescopes at $D=100$~km.

Another important distinction between the TNOs and the Trojans/MBAs is that the survey duration $T$ is a small fraction of the orbital period.  The TNOs are essentially in inertial motion and the deflection due to any tidal force will grow quadratically in time, whereas in the rest frames of the MBAs and Trojans, the applied tidal force has been rotating.  A Planet X at 800~AU will cause a deflection of $\approx180$~km over 10 years, far larger than the signals applied to the MBAs.

\begin{figure}
  \centering
  \includegraphics[width=0.49\textwidth]{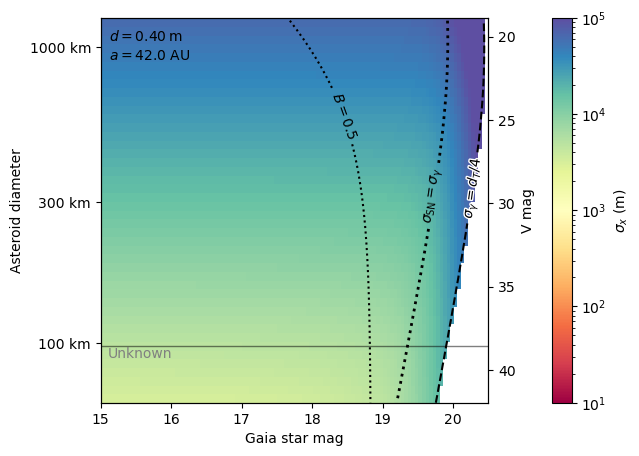}
  \includegraphics[width=0.49\textwidth]{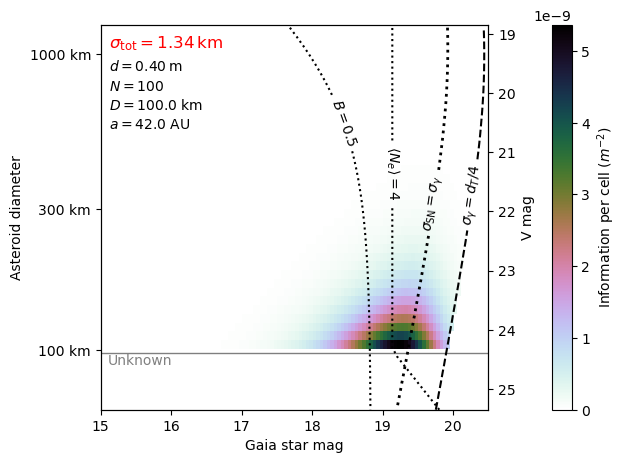}
  \caption{Summaries of the per-occultation measurement errors (left) and total information (right) for an occultation survey targeting TNOs.  All curves have the same meaning as for the MBA case in Figure~\ref{fig:simpleMBA}, except that the color scales and axis limits are changed.  These plots assume $N=100$ telescopes spaced $D=100$~km apart, a much larger array than considered for MBAs and TNOs in order to obtain sufficient occultations per TNO. Further commentary is in Section~\ref{sec:simpleTNO}.}
\label{fig:simpleTNO}
\end{figure}

Figures~\ref{fig:simpleTNO} diagnoses a scenario where $d_T=0.4$~m telescopes (again, nearly optimal) are arrayed at 100~km intervals, spanning 10,000~km---a global-scale effort.   At this point we can see that the TNOs attain $\langle N_e \rangle=4$ events for $m_G\approx19.$  Most of the information is coming from fainter Gaia stars, and background counts produce most of the noise.  The array does, however, attain $\sigma_{\rm tot}=1.3$~km, which is physically much larger than the MBA and Trojan precision levels.  The much larger tidal effects on the TNOs, however, potentially allow for much stronger constraints on unknown masses.

In some sense, a TNO array would be technically easier, in that the events are much slower and rarer.  Our cost formula would describe this as an \$8M array, but certainly operational costs would become much larger for a globe-spanning array than for an array that can be reached by truck on day trips from one or two maintainence facilities.  The RECON network \citep{recon} has deployed 64 $d_T=28$~cm telescopes spanning a $\approx2000$~km North-South path from Kelowna, British Columbia in Canada to Yuma, Arizona in the USA. The RECON stations are operated by local volunteers on an occasional basis, with the primary goal of measuring shapes and sizes of TNOs and Trojans.  While these are not full-time, autonomous stations such as we propose, RECON is a useful demonstration that useful sites at 50--100~km separation can be secured for telescopes.

\section{Simulations}
\label{sec:sims}
Given that the previous Section's rough calculations of the sizes of gravitational perturbations measurable from a feasible occultation array suggest that secure detections of the tidal forces of an $M_X=5M_\oplus,$ 800~AU planet should be possible, we conduct more detailed simulations of such an experiment.  Rather than use $\sigma_{\rm tot}$ as a measure of effectiveness, the simulations use a Fisher-matrix method to estimate the uncertainties in the tidal field itself, after marginalization over all other unknown parameters of solar system dynamics, including the initial state vectors of all test bodies and gravitating bodies.  Using full distributions of MBA and Trojan orbits and sizes, and the real distribution of Gaia stars on the sky, the simulation realizes a full population of occultation events and calculates the positional uncertainties---both along-track and cross-track---expected to be obtained for each event given an observing scenario.

This methodology enables more accurate treatment than the parametric models for many effects, starting with fully realistic distributions of test-body orbits and sizes, the true dynamical perturbations a tidal force would create, and degeneracies of tidal forces with other parameters of the solar system model and non-gravitational forces.  Other effects treated more accurately here include degradation of measurement errors with airmass and moonlight; observability of events given twilight times, weather fractions, and airmass limits; gains in accuracy from multi-chord occultations; and the uneven distribution of Gaia stars on the sky.

We do not include TNOs in this analysis due to the global scale required for a TNO survey to be used for measuring tidal forces. We do, however, predict how many  TNO occultation events would be measurable from a MBA/Trojan-focused array. These observations, though insufficient to provide constraints on Planet X, would increase the number of detailed TNO shapes, sizes, and albedos known to us by $>100\!\times.$

\subsection{Event simulation}

Our survey simulation generates a random sample of occultation events based on the stellar density throughout the path of the target asteroids. We compute a pixelized sky map of the local stellar density, in bins of stellar color and magnitude, from the Gaia DR3 catalog. This allows us to quantify the expected events and their uncertainties without the precise knowledge of minor-body orbits and telescope locations that would be needed to predict each and every event.

A random subset of $\mathcal{O}(10^3)$ orbits is selected from those available in the Minor Planet Center database for each population---MBAs and Jovian Trojans.  The full population of occulters is generated by drawing an absolute magnitude from the expected magnitude distribution of the population within LSST discoverability limit, following estimates by \citet{LSST_science_book}.  A diameter is then computed from the absolute magnitude and albedo through the known expression \citep{HARRIS1997450}
\begin{equation}
\label{harriseq}
\log d_T = 3.1236 -0.5\log q_T - 0.2H.
\end{equation}
To capture the known bimodality in the albedo distribution of MBAs, we randomly select albedo values from the distribution presented by \citet{Murray_2023}. For the Trojans, we assign $q_T = 0.12$ to the entire population, as this is the average for small Jovian Trojans reported by \citet{Yanga2009AJ....138..240F}.

Each simulated target asteroid is then assigned one of the random MPC orbits.  For each 24-hour period of the simulated survey, the asteroid's position in the pixelized Gaia star map is found, and  an occultation rate $n_{occ}$ is assigned according to the length of the asteroid arc $\Delta\alpha$ and local stellar density $n_G$ as
\begin{equation}
n_{occ} = n_G\frac{ND}{D_T}\Delta\alpha,
\end{equation}
where $D_T$ is the Earth-asteroid distance, and angles are expressed in radians. Occultations are drawn from Poisson distributions around the computed rate $n_{occ}$.

Generated occultations are then assigned a random timestamp within the given 24-hour period, and unobservable events are filtered out. For this purpose, latitude is assumed to be $35\degree$N, a rough estimate if the array were to be built in southwest US. Events are limited to night time by selection of those which occur when the sun is lower than $-18\degree$ altitude. Furthermore, events that happen at airmass $> X_{\rm max}=2.3$ are discarded, as well as those closer than $30\degree$ from a more-than-half illuminated Moon. Finally, we discard a random $\approx30\%$ of the nights as having high cloud coverage. 

\subsection{Sources of error}

The expected astrometric error for a single occultation event is a combination of several uncertainties that require careful consideration. Already mentioned in Section \ref{meas_err_1} are the shape noise ($\sigma_{\rm SN}$), photon noise ($\sigma_\gamma$), and Gaia astrometric uncertainties $\sigma_G.$ While the latter is fixed for any given star, the two former ones require further modeling when performing the simulation. Additional sources of error include telescope location and timing of events, as well as angular sizes of stars, since they are assumed as point sources when computing photon noise estimates.

\subsubsection{Shape noise}
\label{sec:shapenoise}

In order to estimate the magnitude of shape noise errors, we simulate occultation data for a population of randomly oriented objects created from shapes found in the 3D Asteroid Catalogue\footnote{\url{https://3d-asteroids.space/asteroids/}}. We then attempt to predict the center of mass of each object, in the direction perpendicular to the array, as the average of occultation chord centers weighted by the length of the chords. For single chord events, the RMS error is $\sigma_{SN} = 0.1 (d_T/2)$; for two or more chords, it approaches $\sigma_{SN} = 0.04 (d_T/2)$.

In the direction of the array, assumed to be North-South, the center of mass location is estimated from the fixed locations of the telescopes involved in the event, being limited by the distance between telescopes or object diameter whichever is smaller. The RMS is therefore expected to be $\sigma\approx{\rm min}(d_T,D)/\sqrt{12}.$

Additional noise will be generated if a binary system is mistakenly assumed to be a single object, because only one member of the system is detected in occultation. In this case, our best center of mass prediction will be offset by a factor dependent on the binary mass ratio and separation. Near-Earth orbit observations indicate a binary fraction of ~15\% \citep{Pravec2006}. Assuming a similar figure for the MBAs, we look at the distribution of mass ratios and separations of known km-sized MBA binaries and we find that most have center of mass offsets smaller or comparable to our nominal shape noise error. We, therefore, do not model this additional error, and introduce the following caveat to our results: ideally, the survey should extend until enough detections of each target are made that additional scatter caused by a binary companion could be statistically inferred.

\subsubsection{Flux and photon noise}
Photon noise error depends on the star's apparent magnitude, as well as background sky brightness. For each occultation event, star magnitude and color are randomly drawn from the color and magnitude distributions previously assigned to the event's sky location. Sky brightness as a function of Moon phase is estimated as specified in Appendix~\ref{sec:diffraction}. When the Moon is below the horizon, the dark sky value is assumed. We further include a linear scaling of background flux with airmass $X$, as well as seeing proportional to $X^{3/5}$. Both relations are in agreement with seeing and sky brightness measurements by the \textit{Dark Energy Survey} \citep{DESschedule}.

Equipped with stellar magnitudes and sky background values, the fluxes before and during the occultation events are computed, and thus, photon noise error for each chord is estimated. The number of chords is drawn based on its probability for each given asteroid size and assigned distance between telescopes. With this draw, occultations that slip between telescopes are also discarded (when zero chords are assigned to them). Having multiple chords reduces the expected photon noise, and the shape noise as described above, but uncertainty propagated from the Gaia star's positional error is independent of the number of chords.

\subsubsection{Telescope location and timing}
Regarding the required accuracy in the timing of events, a clock error of 1~millisecond induces an $\approx 30$~m error in Earth's position, and hence that of the asteroid, which is below the estimated shape noise for our target populations. Similarly it is very easy to conduct geodesy to map each telescope's location on Earth to a level far below our estimated single-event uncertainty $\sigma_x.$ The more difficult challenge, however, may be to limit systematic errors (\ie, correlated among all asteroids) in timing and geodesy so that they are below the combined uncertainty for all our survey data of $\sigma_{\rm tot} \sim 4$~cm found in 
 Section \ref{sec:simpleMBA}. This requires that systematic timing errors be kept below 1 microsecond, and that systematic positioning errors on the stations' mean position be at most at the cm level.  Both should be achievable with GPS devices.

\subsubsection{Stellar angular diameter}
The finite stellar angular diameter is not taken into account in our modeling of diffraction.  If the stellar diameter approaches or exceeds the Fresnel angular scale of the occulter, it will broaden the ingress/egress light curves, and can also decrease the shadow depth, so we must consider it as an additional source of measurement error. We estimate the magnitude of this effect by generating temperature estimates for occulted stars, as follows. For every generated event, star magnitude and color is assigned from Gaia catalog distributions at the given sky location. A temperature value is assigned for each color bin, corresponding to the median temperature of Gaia stars within that bin that have a temperature estimate in the catalog. We then use the magnitude and luminosity relations
\begin{equation}
\label{magnitude}
M = m - 5\log{(d/10\,\text{pc})} - A,
\end{equation}
\begin{equation}
L = L_\odot 10^{-0.4(M + B_{\text{cor}}(T) -M_{\odot})},
\end{equation}
where $A$ is the extinction coefficient and $B_{\text{cor}}(T)$ is the bolometric correction, 
and the Stefan-Boltzmann law,
\begin{equation}
L = \sigma T^4\pi R^2,
\end{equation}
to get the angular diameter
\begin{equation}
\label{diam}
\alpha = 2R/d
\end{equation}
Notice that the result does not depend on the distance (the dependence on $d$ from \eqref{magnitude} is cancelled in \eqref{diam}). Extinction values are estimated based on an average for stars on the given pixelized cell. We use the bolometric correction model for the Gaia G filter from \citet{Andrae2018} for $T < 8000K$, and the LTE+NLTE (non-local thermodynamic equilibrium) statistical model from \citet{Pedersen2020} for $T > 10000K$. The $8000K < T < 10000K$ gap was filled with a linear model that would keep B(T) continuous at $T = 8000K$ and $T = 10000K$.
For most occulted stars, the resultant angular diameter is smaller than astrometric errors from other sources, so stellar angular diameter is not believed to have a significant effect in our results. To be conservative, however, we add the stellar angular diameter in quadrature to our other measurement uncertainties when the star is large enough for this to be relevant.

\subsection{Dynamical model}
We now shift our attention to the propagation of astrometric  measurement errors into uncertainties on external tidal forces by creating the following dynamical model of the solar system. As gravitationally active bodies, we include the Sun, the eight known major planets, plus Pluto and the 343 largest main belt asteroids considered as active bodies by \citet{folkner2014planetary}. The major planets have state vectors and masses as free parameters. The Sun has a free parameter for its mass and its oblateness $J_2,$ but a fixed state vector, in order to remove the coordinate reference frame degeneracy from the system. Pluto and the active asteroids also have fixed state vectors and free masses, since their low masses make it unimportant to know exact state vectors.
 We also model the gravitational potential of the remaining main belt asteroids as arising from two circular rings of mass at $r=2.06$ AU and $r=3.27$ AU respectively, and the azimuthally symmetric portion of the Kuiper Belt with two more rings at $r=39.5$~AU and $r=43$~AU respectively. 
 %We include Pluto at a fixed orbit with free mass, since no observation of Pluto is present in our data set and its low mass ensures us perturbations to its position will not affect much the way Pluto itself perturbs the other bodies. 

There are two relevant non-gravitational forces that act significantly on test particles (the observed MBAs and Trojans)---solar radiation pressure (SRP) and the Yarkovsky effect. The accelerations attributable to these are parallel and perpendicular, respectively, to the sun-asteroid vector, lying in the orbital plane,\footnote{There is a less important component perpendicular to the orbital plane.} and each is inversely proportional to the body diameter, such that
\begin{equation}
    a_{\text{Y,SRP}} = k_{\text{Y,SRP}} \frac{a_\odot}{d_T},
\end{equation}
where $a_\odot$ is the solar gravitational acceleration on the body. 

In our previous paper \citep{Gomes_2023}, we modeled each of the $k$ parameters as global scaling factors. We improve this model by allowing each asteroid to have a free Yarkovsky tangential acceleration parameter. 
This change is required because our predicted per-object center-of-mass uncertainties are smaller than expected displacements from the Yarkovsky force, the latter being of the order of a few kilometers for most MBA targets during a 10 year arc. Therefore, the variance in Yarkovsky displacements due to different asteroid shapes and albedos will exceed the other measurement errors.
 Adding individual Yarkovsky accelerations for each asteroid as constrainable parameters solves this issue, with the only downside that each orbit solution needs an extra data point before it starts being useful for tidal field information: four initial occultations are needed instead of three. On the positive side, our occultation survey would successfully detect this Yarkovsky acceleration parameter on most of our targets, an achievement that otherwise would require many decades of observations \citep{Hung_2023}. For SRP, which has a smaller contribution to orbital displacement, we keep a single global parameter $k_{\text{SRP}}.$

The final element of the dynamical model is a set of 5 point masses at $400$ AU and different fixed positions. \citet{Gomes_2023} show that a Planet X tidal field at any location in the sky can be expressed as a linear combination of the tidal fields from these 5 putative masses. Hence forecasting the covariance matrix of these 5 mass parameters is sufficient to place bounds on the mass of a Planet X in any chosen sky direction.  The quadrupole moment of the 
Kuiper belt mass distribution generates a gravitational field on the Trojans and MBAs that is indistinguishable from a Planet X tidal field.  We therefore add to the $5\times5$ covariance matrix an additional covariance arising from the positions and mass uncertainties 
%
%gravitational signal from large Kuiper Belt objects and assymetries in the Kuiper Belt, which could be degenerate with the signal from a Planet X, is considered by adding to our final covariance estimate a Kuiper Belt floor covariance matrix, computed from the 
of the largest individual TNOs, as well as quadrupole fields from the remaining TNO population, as modeled by the CFEPS project \citep{Kavelaars_2009,Petit_2011,Gladman_2012}. This calculation is described in more detail on our previous work \citep{Gomes_2023}.  The resultant ``Kuiper Belt object (KBO) floor" is dominated by the mass uncertainties of the largest moonless TNOs.

We implement the dynamical model by running the \texttt{REBOUND} $N$-body integrator package \citep{Rein2016} to get variational derivatives of the planets' and test bodies' positions with respect to our model parameters. We start from a base simulation with the Sun and eight planets, to which the massive MBAs are also added. Test particles are included at the initial position of the surveyed MBAs and/or Trojans. We compute variational derivatives, with respect to the active-body parameters and the state vectors of the passive bodies, for the whole time range encompassed by existing historical data on the planets (1965 forward) up to the end of our proposed occultation survey. This is done performing two simulations that start at MJD 60000.0, which integrate respectively forward and backward in time. These simulations are integrated with the \texttt{ias15} adaptive integrator, parallelized for 10 subsets of passive bodies and 30 subsets of main belt asteroids. The small inconsistency between orbits predicted by each simulation, due to the presence of a different subset of massive MBAs, does not significantly affect our forecasts, since they depend exclusively on the variational derivatives, not on the state vectors.

The remaining model elements---ring potentials, solar $J_2$, the mass of Pluto, and tangential acceleration due to Yarkovsky effect---are added one by one to the simulations with two slightly different parameter values, so that their derivatives can be obtained through finite differences. To implement the ring potentials and non-gravitational forces we make use of the \texttt{REBOUNDx} package \citep{REBOUNDx}, as described in \citet{Gomes_2023}. Derivatives with respect to the solar radiation pressure parameter are obtained by rescaling the derivatives with respect to the solar mass.

\subsection{Fisher uncertainties}
Once equipped with our list of simulated observations and their uncertainties, as well as derivatives of model parameters with respect to the observables, we compute the Fisher Matrix to estimate lower bounds on the uncertainties $\sigma_M$ in the mass of a hypothetical Planet X at 400~AU. We follow the method delineated in \citet{Gomes_2023}: first, a Fisher Matrix is  generated for the set of existing  observations of the major planets with respect to our global model parameters, namely the masses and state vectors of the planets, masses of $\approx300$ asteroids, collective mass rings for smaller MBAs and TNOs, and the tidal forces of interest. The observations we use for this step are the same ones enumerated in the aforementioned paper, and they consist of astrometry and radio ranging measurements. Next, for each small-body test particle, the Fisher matrix is augmented with rows and columns for its 
initial-state-vector parameters (and, in our current case, a Yarkovsky effect strength), and then the Fisher information from its simulated observational data is added in. We can then marginalize the Fisher matrix over this body's parameters. The Fisher matrix thus remains fixed in size even as we accumulate constraints from millions of test bodies.
Finally, we invert the total Fisher matrix to form a covariance matrix, and extract the $5 \times 5$ sub-matrix corresponding to the five putative distant point masses. To the resultant covariance matrix, we add the previously mentioned Kuiper Belt floor covariance. Then, for any given position in the sky, a transformation of the covariance matrix, as detailed in Appendix A of \citet{Gomes_2023}, yields us the lower bound uncertainty $\sigma_M$ on the mass of a Planet X in any chosen direction from the Sun.

We produce three distinct forecasts: first using only MBAs as test particles; then using only Trojans; and then for the scenario of a joint survey of both populations.

\section{Simulated constraints}
\label{sec:results}

We consider a grid of possible telescope arrays that vary between $N=50$ and $N=400$ telescopes, as well as telescope diameters between $d=20$~cm and $d=70$~cm. For each array configuration, we measure the constraining ability of the array by predicting $\sigma_M$ of Planet X on a full sky grid of possible point mass positions, and selecting the 90th percentile value; \ie, the $\sigma_M$ that can be achieved or surpassed over 90\% of the sky in that particular survey configuration.

\begin{figure}
  \centering
  \includegraphics[width=0.49\textwidth]{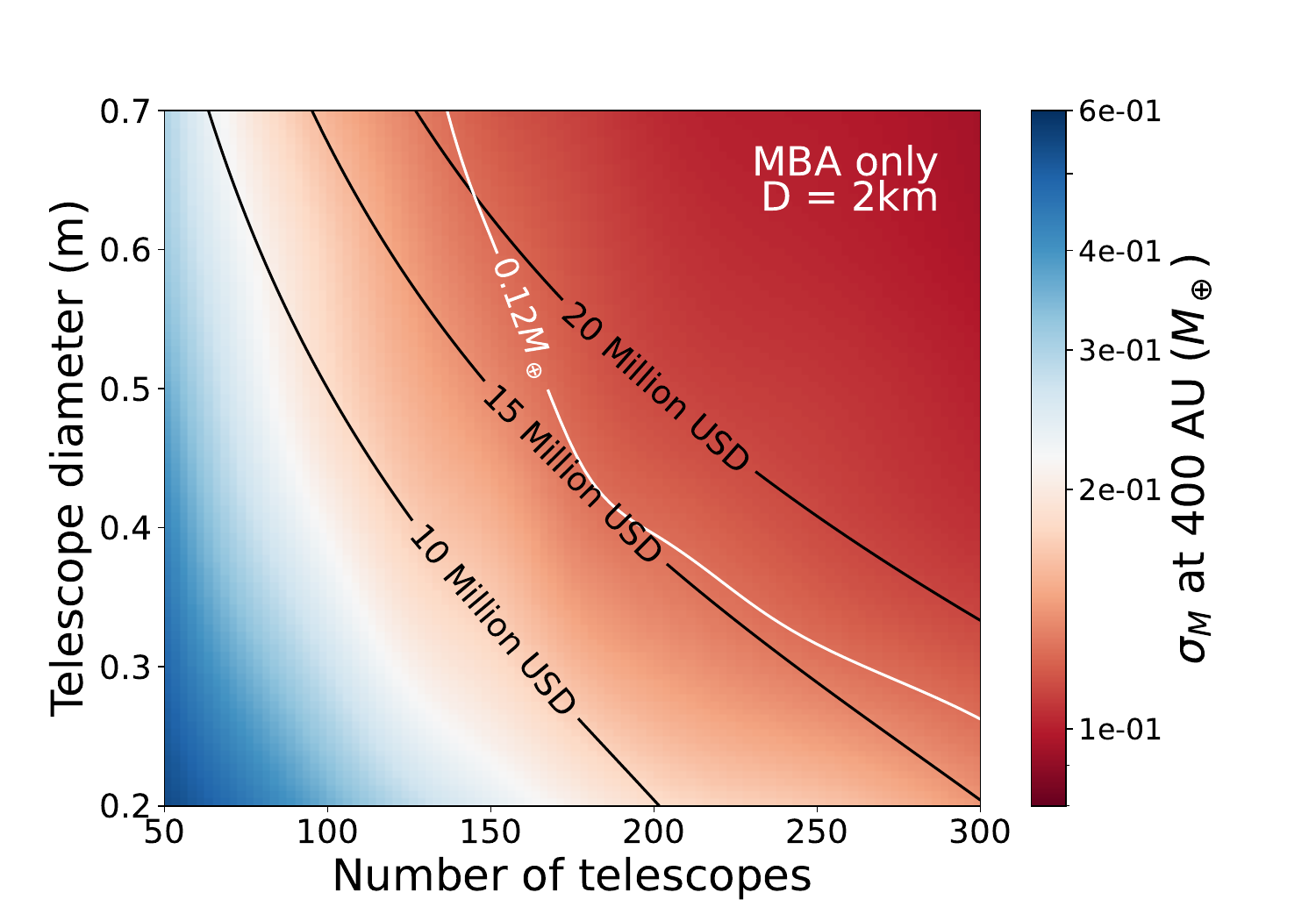}
  \includegraphics[width=0.49\textwidth]{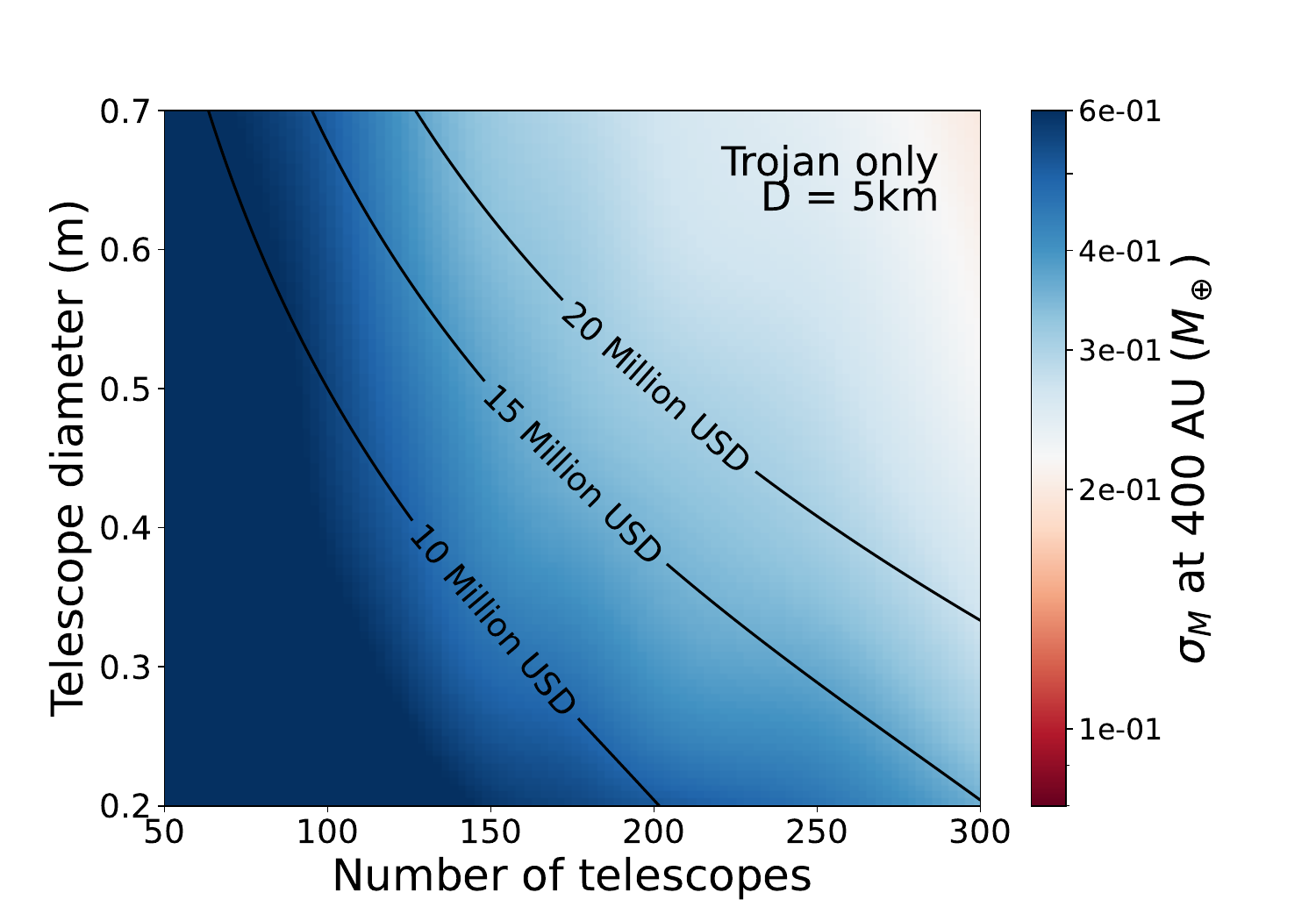}
  \includegraphics[width=0.49\textwidth]{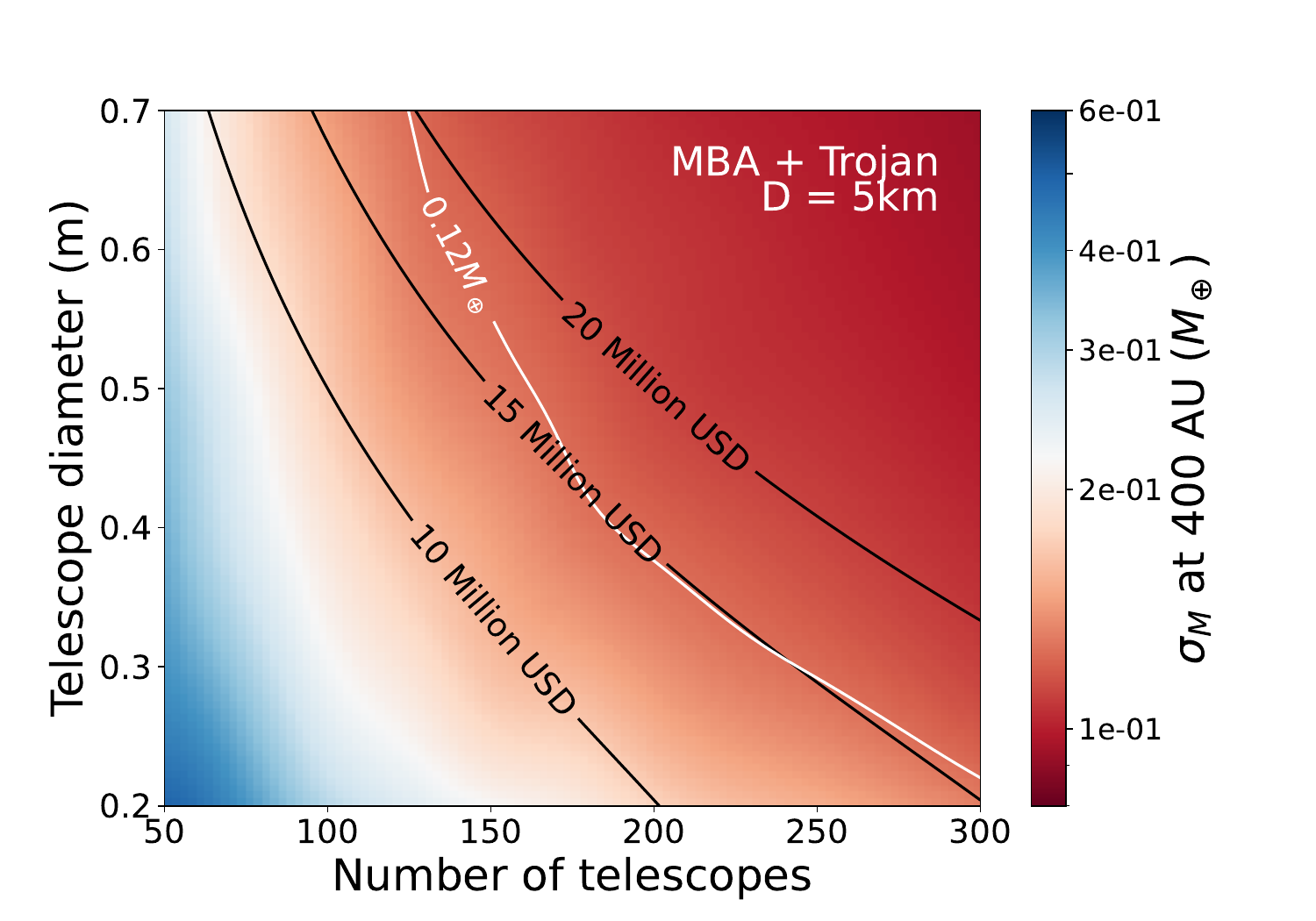}  
  \caption{Value of $\sigma_M$ under which a point source at 400AU would be constrained in 90\% of the sky. Black contours are lines of equal cost. The white contour at $\sigma_M = 0.125M_\oplus$ corresponds to a 5$\sigma$ detection at 800AU. Top line: Left panel shows results for MBA survey with $2~km$ distance between telescopes. Right panel shows Jovian Trojan survey, with $5~km$ distance between telescopes. Bottom line: Combined MBA and Trojan survey, with $5~km$ distance between telescopes.}
\label{fig:sim_constraints}
\end{figure}

Three scenarios are presented in Figure \ref{fig:sim_constraints}. The first one (top left), assumes the array would be exclusively devoted to MBA occultations. In this case, the distance between telescopes is set at $D=2$~km.  There is no change in $\sigma_M$ at larger $D$, since the majority of occulters are smaller than 2~km and the number of observed occultations becomes independent of $D$.  The second scenario (top right) assumes an exclusive Jovian Trojan survey, and sets the distance between telescopes to $D=5$~km.
Larger $D$ is favored for the Trojan survey, since they are twice as far as MBAs, making the smallest ones detectable by LSST larger than for MBAs. The third scenario (bottom) assumes a joint MBA/Trojan survey, with $D=5$~km again since the Trojans benefit and the MBA observations are not harmed by increased $D$.

\begin{figure}
  \centering
  \includegraphics[width=0.7\textwidth]{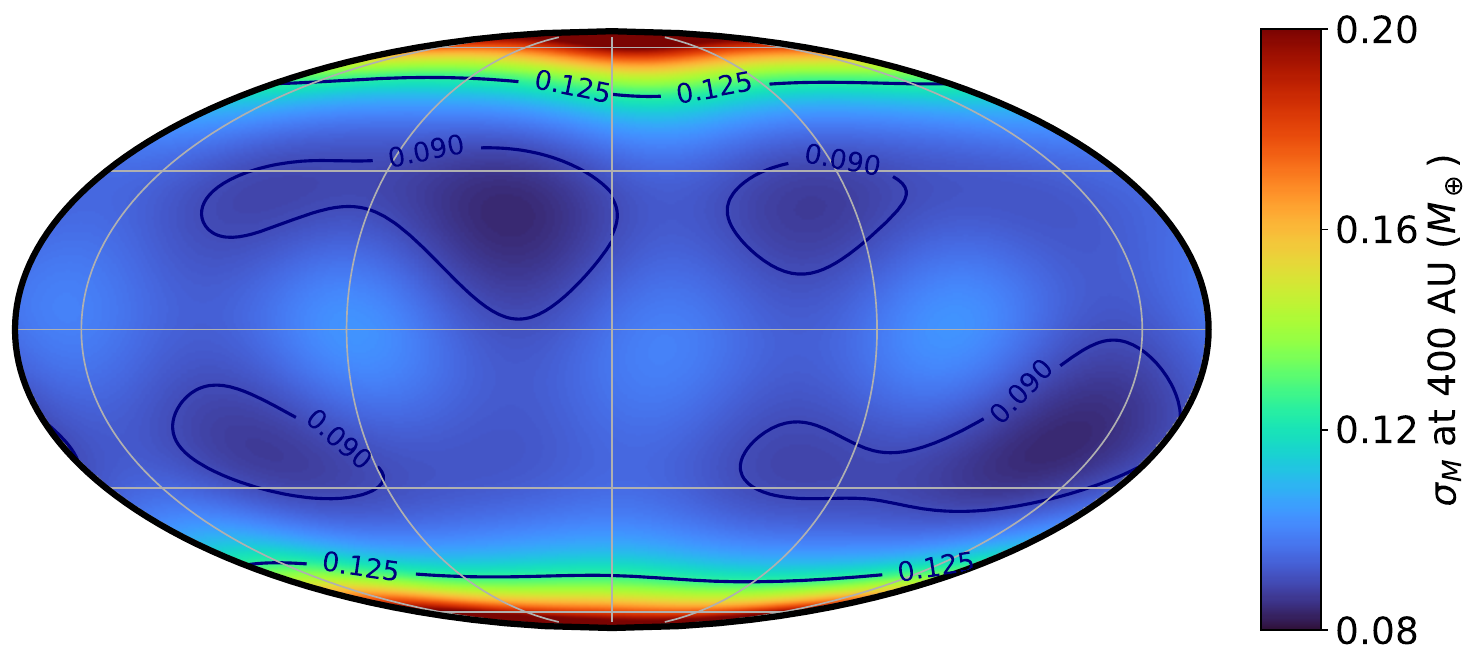}
  \caption{Lower bound on the uncertainty $\sigma_M$ for a point source at 400AU as a function of its location in the sky. We use the Mollweide projection and ecliptic coordinates with the vernal equinox at center.}
\label{fig:skymap}
\end{figure}

The cost of each array configuration is computed according to Eq. \ref{eq:cost}, so that the optimal configuration for any given cost can be easily determined. In all scenarios, the optimal array diameter varies within the 30--40~cm range, and optimal number of telescopes increases with the allowed budget. A $5\sigma$ detection of a $5M_\oplus$ body at 800AU, which is the smallest and most distant hypothesized Planet X invoked to explain alignment anomalies of distant TNOs, is achieved if $\sigma_M=0.125M_\oplus$ (the tidal field of such a planet would be 8 times weaker than an equal mass at 400AU). Figure \ref{fig:sim_constraints} shows us that a cost-optimized array capable of such detection would be slightly above 15 million USD for an MBA-exclusive survey. Trojans would not independently allow such a Planet X constraint, but they do slightly help the MBA result in a joint survey, making the same precision cross below 15 million USD. Optimal configurations in both these scenarios are around $N\sim 200$ telescopes and $d\sim 40$~cm. Figure \ref{fig:skymap} shows the forecasted uncertainty $\sigma_M$ of Planet X as a function of its location in the sky for a joint $N = 200$, $d = 40$~cm scenario.

To assess the feasibility of performing observations at or close to the predicted event frequency, we assume that a single measurement uses $\sim$ 2 minutes of telescope time. The optimal configuration is therefore limited at $\sim$ 6000 events $\times$ telescopes per hour. Our simulations recorded hourly rates of successful observations (on nights with clear sky) varying in the 2000--3000 range. This means that, if exactly one  telescope and time slot is required for each of 3000 successful events, the ``subscription factor'' would be $f\approx0.5$ on the busiest nights, and if the events are randomly distributed over stations and time, a fraction $e^{-f}-(1-f)\approx0.1$ would overlap another event and be lost.  If, however, 2 or more time slots are needed for each success, either because the orbit uncertainty spans multiple telescopes, or the targets are small enough to slip between stations, then $f\approx 1$ or more, meaning $\approx40\%$ or more of events would be lost to resource conflicts.
This is most concerning at the initial stage of the survey, when target ephemerides are limited by LSST astrometry, and we expect a few failures before attaining the first successful observation of each target. Therefore, we would definitely benefit from array design options that  push towards higher event frequency capabilities, and/or faster event cycles than 2~minutes. If the survey, before most objects get their first detection, is ultimately limited by the event load, we might need to extend survey duration to recover our predicted constraints.

\begin{figure}
  \centering
  \includegraphics[width=0.7\textwidth]{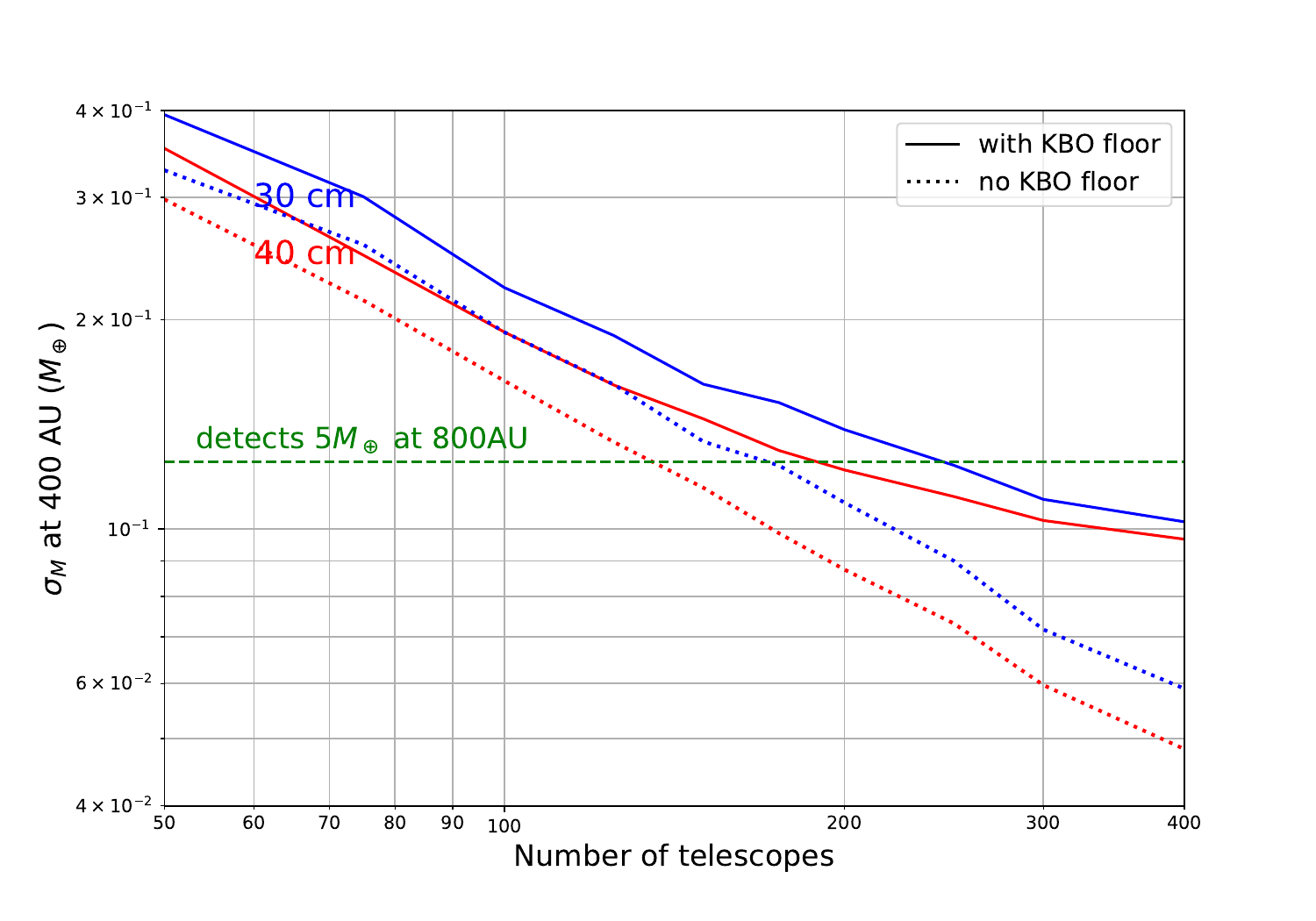}
  \caption{Value of $\sigma_M$ for a point source at 400AU that would be achieved or surpassed for 90\% of possible directions, for a joint MBA/Trojan survey. Blue line assumes 30cm telescopes; red line, 40cm. Solid lines include fundamental degeneracy of the tidal signal with the uncertainties of large TNO masses and asymmetries in the Kuiper Belt (KBO floor). Dotted lines show the predicted uncertainty disregarding this limitation.}
\label{fig:sim_constraints-1d}
\end{figure}

\begin{figure}
  \centering
  \includegraphics[width=0.7\textwidth]{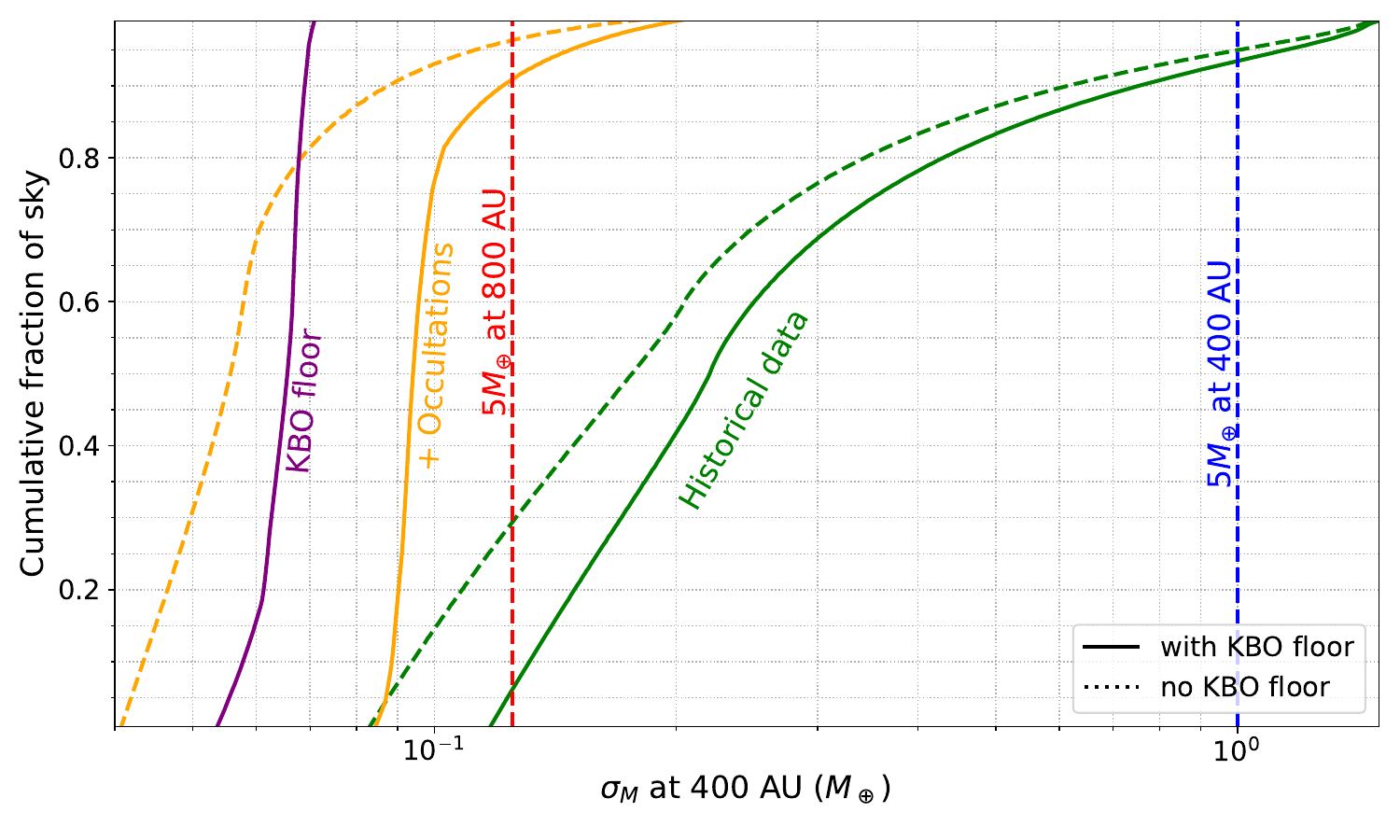}
  \caption{Fraction of the sky where the lower bound on $\sigma_M$ lies below a given value. Dashed lines disregard the degeneracy of a Planet X tidal field with Kuiper Belt tidal field uncertainties.}
\label{fig:cumul}
\end{figure}

In Figure \ref{fig:sim_constraints-1d}, we see the decrease of $\sigma_M$ for the joint survey as a function of the number of telescopes for  both the $d=30$~cm and $d=40$~cm scenarios. Additional dotted lines show the constraints that would be attainable were there no degeneracy with Kuiper belt tidal fields; \ie, computed out of the original covariance matrix, without a KBO floor addition. We find a strong gain in sensitivity for additional investment in $N$, with $\sigma_M \propto 1/N$ approximately---much better than the typical square-root relation.
With the assumed  Kuiper Belt floor, however, there is little gain in $\sigma_M$ for $N\gtrsim200$ and the associated cost increase---although one would be measuring the KBO mass distribution, and other science benefits of the project beyond measuring $\sigma_M$ would improve.
 Finally, Figure \ref{fig:cumul} shows the cumulative fraction of the sky where uncertainty $\sigma_M$ falls below a certain level, in two different scenarios: using exclusively historical data, and adding occultations assuming the optimal array to historical data. 

We now consider the optimal $\sim 15$ million USD array ($N\sim 200$, $d\sim 40$~cm) and ask how effective such a configuration would be towards detecting TNO occultations. We use sample orbits from the CFEPS L7 population \citep{Kavelaars_2009,Petit_2011,Gladman_2012} and find that, for a 10 year survey, the array measures one occultation by $6\%$ of the simulated objects (multiple occultations for a TNO are rare). For the assumed number of 30,000 LSST-detectable TNOs, a $6\%$ rate would mean 1800 distinct TNO shapes. Though such data will not provide constraints on orbital perturbations, a single occultation is still useful.  A single occultation of these $d_T\gtrsim100$~km objects with chords recorded at 5~km intervals will return a detailed projected size and shape measurement of the TNO, leading to a better understanding of the size, albedo, and contact binary rate for a substantial number of TNOs in all dynamical classes. It will also yield more accurate predictions of subsequent occultation events, which could potentially be observed by other occultation campaigns.


\section{Conclusions}
We present the expected outcome of a proposed dedicated occultation array with $\mathcal{O}(100)$ telescopes, cost-optimized to constrain minor planets' orbits for detection of yet-undiscovered tidal fields. An initial parametric analysis shows comparable results for arrays targeting MBAs and Jovian Trojans, with the $5 \times$ weaker spatial precision for Trojans compensated by the presence of larger tidal perturbations on their orbits. Combination of data from all targets would provide cm-level sensitivity to tidal perturbations. A TNO-focused array would require different logistics, due to the slower apparent path of these objects on the sky and smaller known population. To achieve the three required observations that initially constrain the orbit, such an array would need to be spread out across a $\sim 10,000$~km baseline between telescopes, turning the project into a global effort.

Survey simulations provide further insight into the MBA and Trojan array outcomes, by considering stochastic factors that affect observability of each independent event. The simulations yield Fisher-matrix estimates of the 
$\sigma_M$ of the mass of any undiscovered Planet X lurking in the solar system, including the effects of degeneracies with other sources of orbital perturbation. We find that MBA surveys achieve $2-3 \times$ lower uncertainties than the Trojan ones, which is not as optimistic for the Trojans as our parametric analysis. We suspect that this results from the lower typical number of occultations observed for an individual Trojan than for an MBA, which makes it more difficult to disambiguate a tidal force from other dynamical perturbations.
%This difference is due to the fact that simulations are actually probing degeneracies between Planet X signatures and other possible perturbers. Since Trojans get a smaller number of occultations on average, this difference is seen more drastically on them.  
We consider which scenarios could yield a $5\sigma$ detection of a $5M_\oplus$ Planet X at 800AU on at least $90\%$ of the sky. The lowest-cost option is a joint MBA/Trojan survey with $N\approx 200$ telescopes spaced at $D=5$~km (\ie\ a 1000~km array), each with diameter $d\sim 40$~cm, and a very crude construction cost estimate of $C\approx$15 million USD.

When compared with the previous suggestion of constraining a Planet X signal with LSST astrometry of Jovian Trojans \citep{Gomes_2023}, an occultation survey is much more promising, mainly because it gathers information from the faintest end of the asteroid size distribution. LSST astrometry is assumed, however, as a pre-requisite for accurate occultation predictions. 

The constraint on missing mass in the solar system would not be the only science benefit of the small-telescope linear array.  
Our survey would allow the study of other perturbations to asteroid orbits, constraining Yarkovsky acceleration parameters and detecting mutual deflections by close asteroid encounters. These data, therefore, would contribute to a better understanding of asteroid densities, albedos, and thermal properties. 
Such an array would also provide detailed shapes for $\sim 1800$ TNOs, characterizing the distribution of sizes, shapes, albedos, and contact-binary rates of TNOs in many different dynamical states.  No other envisioned observation could yield such information for a large sample of pristine products of planetesimal formation.

Characterizing orbital perturbations with the proposed level of accuracy can lead to further discoveries in physics. The hypothesis that primordial black holes might be the constituents of dark matter predicts rates of close encounters between such objects and solar system bodies. \citet{Tung2024} point out that resultant perturbations would carry the particular signals of high perturber velocity ($v \approx 200$~km/s) and non-ecliptic trajectory, and would be distinguishable from other interstellar perturbers in that baryonic  objects of comparable mass would be easily identifiable. Their proposed method of detection requires high-precision astrometry of multiple solar system objects, so that perturber information can be inferred from correlation between perturbations on different orbits---this aligns exactly with the goals of our array.

Sensitivity to cm-scale perturbations out to 5~AU in a joint MBA-Trojan survey would yield constraints on modifications to General Relativity as well.   Looking beyond occultation observations, an array of hundreds of agile robotic telescopes, with a total collecting area equivalent to a $\approx6$~m telescope, could conduct other investigations between scheduled occultation events, such as targeted or blank-sky monitoring for transients.  This would include monitoring of sources that are too bright or fast for LSST or that require cadences that do not fit in the LSST survey plan.

As for the location of the array, we are limited by latitude, since a high-latitude survey would be blind to significant portions of the ecliptic plane below the horizon. We also require a North-South stretch of $\mathcal{O}(1000)$~km with good observing weather, low light pollution, accessible terrain, and, preferentially, some pre-existing infrastructure, namely a road, allowing for easy deployment of telescope stations and subsequent access for repairs and maintenance, and---if no cellular network is in place---collection of physical data storage media. Five candidate locations, selected with these conditions in mind, are: the southwest US, northern Chile, north-central Australia, Namibia/South Africa, and Saudi Arabia.

\begin{acknowledgments}
This work was supported by National Science Foundation grant AST-2205808.  We thank Marc Buie, Matt Lehner and Zachary Murray for helpful discussions.
\end{acknowledgments}

\bibliography{refs}{}

\appendix
\section{Symbol Glossary}
Table 1 displays the main symbols used throughout this paper, their meaning and default value, where applicable.
\startlongtable
 \begin{deluxetable}{clc}
  \label{tab:symbol}
  \tablewidth{0pt}
  \tablecaption{Symbol glossary}
  \tablehead{ \colhead{Symbol} & \colhead{Meaning} & \colhead{Default value} }
  \startdata
  \cutinhead{Observation specifications}
  $d$ & Telescope diameter & \nodata \\
  $D$ & Telescope array spacing & 2~km \\
  $N$ & Number of telescopes in array & \nodata \\
  $C_s$ & Fixed cost per observing station & \$40k \\
  $C_t$ & Cost per $d=0.5$~m telescope, scaling with area & \$60k \\
  $T$ & Survey duration & 10 yr \\
  $\lambda_0$ & Central wavelength of passband & 600~nm \\
  $\eta$ & Quantum efficiency of atmosphere/telescope/detector & 0.5 \\
  $f_{\rm duty}$ & Duty cycle (fraction of occultations observed) & 0.21 \\
  $X$ & Airmass of observation & \nodata \\
  $X_{\rm max}$ & Maximum observable airmass & 2.3 \\
  $\beta_{\rm min}$ &  Minimum observable solar elongation & 50\arcdeg \\
  $\theta_{\rm ap}$ & Angular radius of photometry aperture & 1\arcsec \\
\cutinhead{Occulter specifications}
$a_T$ & Semi-major axis of occulter (target) & \nodata \\
$d_T$ & Diameter of occulter & \nodata \\
$q_T$ & Geometric albedo of occulter & 0.10 \\
$D_T$ & Distance to occulter & $a_T$ \\
$v_\perp$ & Relative transverse velocity of observer and occulter & $v_\oplus=3\times10^4$~m/s \\
$m_{\rm lim}$ & Faint limit of known (LSST) occulters & $m_r=24.4$ \\
$N_T(>d_T), dN_T/dd_T$ & Cumulative, differential counts of occulters & \nodata \\
\cutinhead{Source star properties}
$n_G(<m), dn_G/dm$ & Cumulative, differential sky densities of usable Gaia stars & \nodata  \\
\cutinhead{Derived quantities}
$F$ & Fresnel length $\sqrt{\lambda_0 D_T/2}$ & \nodata \\
$t_F$ & Fresnel crossing time $F/v_\perp$ & \nodata \\
$t_0$ & Time of chord center & \nodata \\
$\tau$ & Characteristic time scale of occultation & $t_F$ \\
$f(u)$ & Occultation light curve as function of scaled time $u=(t-t_0)/\tau$ & \nodata \\
$n_b$ & Rate of background photocarriers & \nodata \\
$n_\star$ & Unocculted rate of star photocarriers & \nodata \\
$B$ & Background ratio $n_b/n_\star$ & \nodata \\
$m_G^F$ & $G$ magnitude of star producing 4 counts per $t_F$ & \nodata \\
$m_G^B$ & $G$ magnitude of star for which $n_\star=n_b$ & \nodata \\
$\rho$ & Scaled occulter radius $d_t/2F$ & \nodata \\
$\sigma_t$ & Total uncertainty on $t_0$  & \nodata \\
$\sigma_x$ & Total along-track uncertainty on chord center & $v_\perp \sigma_t$ \\
$\sigma_\gamma$ & Contribution to $\sigma_x$ from photon noise & \nodata \\
$\sigma_{\rm SN}$ & Contribution to $\sigma_x$ from occulter shape & $0.05 d_T$ \\
$\sigma_G$ & Contribution to $\sigma_x$ from Gaia position & DR5 spec \\
$N_e$ & Number of observed occultations per asteroid & \nodata \\
\enddata
\end{deluxetable}

\section{Measurement errors on chord midpoints}
\label{sec:diffraction}
A photon-counting aperture photometer will be drawing a Poisson sample of events from an occultation light curve with a time-dependent rate given by
\begin{equation}
  n(t) = n_b + n_\star f\left(\frac{t-t_0}{\tau}\right),
\end{equation}
where $n_b$ and $n_\star$ are the count rates generated by background sources and the unocculted star, respectively.  In this context, ``background'' refers to any source other than the star being occulted, including diffuse zodiacal and atmospheric emission, light reflected from the occulting asteroid, or neighboring stars, all of which can be expected to remain constant over the fraction of a second when the telescope is in ingress or egress of the shadow.  The function $f$ gives the fraction of stellar intensity making it past the occulter, \ie\ the light curve relative to the time of passage $t_0$ of the geometric shadow edge over the telescope, with time specified in units of some time scale $\tau.$  What is $\sigma_t,$ the expected RMS uncertainty in $t_0$ given the data $D,$ where $D$ in this case is a list of arrival times of detected photons?  There is not, to our knowledge, any general analytic form for this ``change detection'' problem.  We construct an approximate formula for $\sigma_t$ based on analytic results for some limiting cases, and results of maximum-likelihood estimators applied to simulated photon streams.

The Cramer-Rao theorem bounds $\sigma_t$ using the Fisher information on $t_0,$ which in this case can be written as
\begin{align}
  \sigma^{-2}_t & \le \left\langle \left( \frac{\partial \log p(D | t_0)}{\partial t_0} \right)^2\right\rangle \\
                & = \int dt\, \frac{ \left[n_\star f^\prime(t/\tau)/\tau\right]^2}{n_b+n_\star f(t/\tau)} \\
                & = \frac{n_\star \tau}{\tau^2} I(B), \label{eq:fisher1} \\
  I(B) & \equiv \int du \frac{\left[f^\prime(u)\right]^2}{B + f(u)} \label{eq:IB} \\
  \Rightarrow \quad \sigma_t & \ge \frac{\tau}{\sqrt{n_\star \tau}} I(B)^{-1/2}. \label{eq:fisher2}
\end{align}
where we have defined $B=n_b/n_\star$ as the background ratio.
In the first equality, we have exploited the fact that the Fisher information in a Poisson distribution with a mean of $\lambda(q)$ for some
parameter $q$ is equal to $(d\lambda/dq)^2 / \lambda,$ and we can divide the light curve into statistically independent time segments of duration $\Delta t$ with expected mean count rates of $\lambda = n(t)\,\Delta t.$

\begin{figure}
  \centering
  \includegraphics[width=0.8\textwidth]{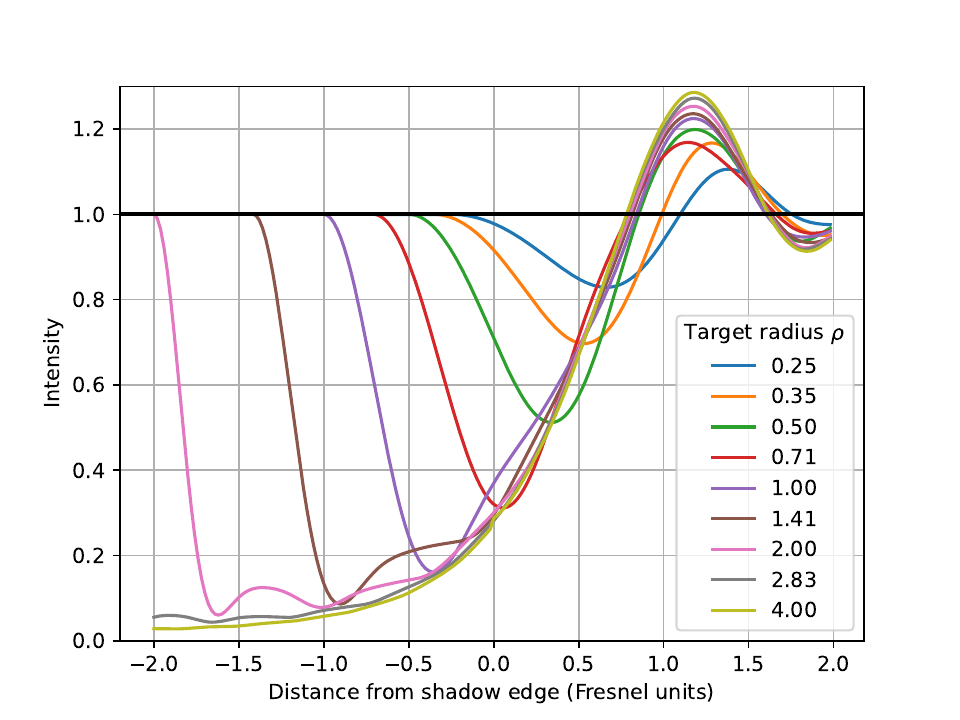}
  \caption{Light curves for the egress portion occultations of a point source by a spherical occulter of radius $\rho F,$ where $F=\sqrt{a\lambda/2}$ is the Fresnel scale.  The horizontal axis can be considered either the distance from the geometric shadow edge (in units of $F$), or the time since the crossing of the geometric shadow edge, in units of $t_F=F/v_\perp.$}
  \label{fig:shadows}
\end{figure}

For the ingress or egress of occultation, the relevant time scale is $t_F=F/v_\perp \approx \sqrt{\frac{a\lambda}{2}}/v_\oplus,$ the time for the Fresnel scale to cross a telescope moving at $v_\perp$ across the line of sight to the target.  \eqq{eq:fisher2} shows that the uncertainty on the start or end time of a chord is the Fresnel time divided by the square root of the number of source photons collected during the Fresnel time.  The $I(B)$ factor encodes dependence on the size/shape of the target and the effect of background.

For a simple model in which $f(u)$ transitions linearly from 0 to 1 over a range $\Delta u=1,$ \ie\ a transition time $\tau$ to complete shadow, the Cramer-Rao inequality yields
\begin{equation}
  \sigma_t \ge \frac{\tau}{\sqrt{n_\star \tau}} \left[ \log(1+1/B)\right]^{-1/2}.
    \label{eq:linearfisher}
\end{equation}
The Cramer-Rao theorem does not prescribe an estimator for the time of ingress/egress, it just bounds any unbiased estimator's performance.  We might expect the Fisher value to be attained when $\sigma_t\ll \tau,$ when the differentials used in the Fisher matrix are applicable.  We tested a straightforward estimator for the time of egress, namely the expectation value $\hat t_0 = \int dt_0\, t_0 {\mathcal L}(D|t_0) /  \int dt_0\,  {\mathcal L}(D|t_0)$.  We find that equality in \eqq{eq:linearfisher} is indeed attained, within $\approx10\%,$  by this estimator when $\sigma_t\lesssim0.5$ and $B\gtrsim 0.02.$  This gives us confidence to use the Fisher estimator for $\sigma_t$ with the more complex $f(u)$ light curves obtained from diffraction, for cases when the uncertainty will be well below the Fresnel time.

To estimate typical $I(B)$ factors for diffractive occultations, we follow \citet{Nihei2007AJ} by using the results of scalar diffraction theory for spherical bodies \citep{sommargren}.  Figure~\ref{fig:shadows} plots the transmitted intensity function $f$ for the egress portion of 
a central chord across a body of radius $\rho F.$ Each light curve is integrated over an octave of wavelength $2\lambda_0/3 < \lambda < 4\lambda_0/3,$ and the horizontal axis is the variable $u=(t-t_0)/\tau,$ with $t_0$ at the geometric shadow edge.  Occulters of all size tend to follow a common curve in the neighborhood of the geometric shadow edge, with smaller bodies essentially having a maximum depth of shadow that truncates the curve.  For each combination of $\rho$ and background factor $B$ we evaluate the integral in \eqq{eq:IB}, using only the time period between the minimum and maximum of the light curves in Figure~\ref{fig:shadows}. This period of time shows a largely monotonic rise, avoiding the central ``Arago spot'' and any oscillatory regions, which are likely to be strongly dependent on the detailed shape of the occulter.

The following functional form reproduces the resulting values of $\sigma_t$ to within $<5\%$ over the relevant range of target radius $\rho$ and all values of background ratio $B$:
\begin{align}
  \sigma_t^{\text{Fisher}} & = 0.77 \frac{\tau}{\sqrt{n_\star \tau}}  \times \left[1+(0.35/\rho)^{5/2}\right] \times \sqrt{1 + m_B(\rho) B}, \\
  m_B(\rho) & = \text{min}( 1+\rho^2, 1.8).
\end{align}              

This result assumes central chords of the sphere.  Numerical integration of the Fisher information for light curves for chords off-center by fraction $b$ of the occulter radius show weak degradation of $\sigma_\gamma,$ by factors $\approx(1-b^2)^{-1/8},$ to within 10\% accuracy, which means that the mean information from random sampling of $b \in [-1,1]$ is only $\approx10\%$ lower than the calculation assuming central chords.

We need next to consider cases in which the Cramer-Rao lower bound is $\gtrsim \tau/2.$ In such cases, occultations of duration $\lesssim 2\tau$ will be essentially undetectable, which means that this case is applicable only to sources with $\rho>1.$  For such objects, the shadow is dark, with $f(u)$ decreasing to $\lesssim0.1.$ Furthermore, if the transition from $f=1$ to $f\approx 0$ is not resolved by the data, then the shape of this transition is unimportant, and we can assume an instanteous drop.  This leaves us in the simple geometric-optics limit, whereby the photon rate drops instantaneously from $n_\star+n_b$ to $n_b$.  Scaling arguments then tell us that we must have $\sigma_t = S(B)/n_\star$ for some function $S$ of the background-to-source flux ratio $B.$  The case of $B=0$ can be treated analytically: the estimator will clearly involve just the time of the first (last) arriving photon for egress (ingress), and one can show that $S(B=0)=1.$  We can also expect $S(B)\propto B$ for $B\gg 1$, from the relation that the signal-to-noise ratio of the dip in flux over time period $t$ is $S/N=n_\star t / \sqrt{n_b t}$.  We find an empirical approximation to $S(B)$ for geometric shadows that matches both expected limits, and is accurate to $\approx10\%$ in between for simulations using our likelihood-based estimator for $t_0$:
\begin{equation}
  \sigma_t^{\text{geom}} = \frac{1}{n_\star} \left[ 1 + \left(\frac{e}{\log(1+1/B)}\right)^2\right]^{1/2}.
\end{equation}

Our final estimator for the photon-noise uncertainty in the position of the center of the chord is to join  $\sigma_t^{\text{Fisher}}$ with $ \sigma_t^{\text{geom}}$ at the point where they cross, which is typically when $\sigma_t\approx \tau/2.$  We increase the Cramer-Rao bound by a factor 1.1, and divide each result by $\sqrt{2}$ to account for the chord midpoint being the average of the ingress and egress times.  Then the time uncertainty is converted to a distance uncertainty with a factor of $v_\perp,$:
\begin{equation}
  \sigma_\gamma = \frac{v_\perp}{\sqrt{2}} \times \max \left[ 1.1\sigma_t^{\text{Fisher}}(\tau=F/v_\perp, \rho, n_\star, B), \sigma_t^{\rm geom}(n_\star, B) \right]
\label{eq:sigmagamma}
\end{equation}
It is also the case that occultations for which $\sigma_\gamma \gtrsim d_T/2$ will not be reliably detected as decrements in stellar photon fluxes, hence offer essentially zero information on the asteroid's orbit.

Examining the above equations as a function of $n_\star,$ or equivalently the source star's magnitude $m_G,$ we find that there are two important inflections in the behavior of $\sigma_\gamma$ at fixed values of $n_b.$ One is the value $n_\star^F$ such that $n_\star^F t_F=4,$ four photons being detected during the ingress/egress ramps.  This is roughly the point at which the Fisher approximation begins to fail, and the Fresnel time is no longer resolved by the data. Assuming a one-octave filter centered at 600~nm, the magnitude in Gaia $G$ band at which this occurs is
\begin{equation}
  m_G^F = 17.6 -2.5\log_{10}\left[ \left(\frac{a}{2.6\,\text{AU}}\right)
  \left(\frac{0.5\,\text{m}}{d}\right)^2
  \left(\frac{0.5}{\eta}\right) \right]
\end{equation}
with $\eta$ being the quantum efficiency (QE) of the atmosphere/telescope/detector combination.    The other important inflection point is when the background rate equals the source rate, $B=1.$  This magnitude $m_G^B$ is dependent only on the filter bandpass, sky characteristics at the site, and the radius $\theta_{\rm ap}$ of the photometric aperture.  We estimate this quantity using on-sky data for similar filters on the Dark Energy Camera on the Blanco telescope at Cerro Tololo\footnote{Extracted from the \href{https://noirlab.edu/science/documents/scidoc0493}{DECam exposure time calculator}.} and from the Megacam on the Canada-France-Hawaii Telescope\footnote{S. Gwyn, private communication.}.  These estimates are in rough agreement with each other; we take the higher-background result as a conservative choice, keeping in mind that a real occultation array would need to be at lower altitude with higher background than these mountaintop observatories.  For dark skies at zenith, a typical value for star-sky equality is
\begin{equation}
  m_G^B = 19.6 - 5\log_{10} \left(\frac{\theta_{\rm ap}}{1\arcsec}\right).
\end{equation}
This is faintward of $m_G^F$ for MBA occulters and a 0.5~m telescope.  Figure~\ref{fig:sigx} plots the measurement error $\sigma_\gamma$ of a chord midpoint position vs $m_G$ for a nominal MBA with diameter well above $F\approx300$~m, and a particular choice of telescope diameter $d$.
For stars brighter than the inflection points, $\sigma_\gamma \propto f_\star^{-1/2}$, but faintward of the inflections, stars lose power more rapidly, with $\sigma_\gamma \propto f_\star^{-3/2}.$  There is thus little to be gained by observing occultations of stars fainter than these inflection points.

We have not addressed the effect of scintillation on the chord detection.  The time scale $t_F=11\,{\rm ms})\times (a/2.6\,\text{AU})^{1/2}$ for the light curve is only slightly shorter than the scintillation time scale implied by a 10~m/s wind crossing a telescope of size $d=0.4$~m, so there will be some scintillation effects on the relevant portion of the light curve. Figure~\ref{fig:shadows}, however, suggests that occultations for $\rho>0.5$ have a depth greater than 50\%, which is substantially larger than the amplitude of scintillation fluctuations. So this effect deserves further study, but is unlikely to change results dramatically.

\end{document}